\documentclass[
reprint,
superscriptaddress,
 amsmath,amssymb,
prx,
]{revtex4-2}

\usepackage{float}
\usepackage{graphicx}
\usepackage{xcolor}
\usepackage{amsmath,amssymb,amsthm,amsfonts}
\usepackage{bbm}
\usepackage{bm}
\renewcommand{\vec }{\bm}
\renewcommand{\mathbf}{\bm}
\usepackage{enumerate}
\usepackage{ulem}
\usepackage{dsfont}
\usepackage{makecell}
\usepackage{threeparttable}

\begin{document}
\title{Assessing the atomic moment picture of spin dynamics: the perspective of \textit{ab initio} magnon wavefunction
}

\author{Yihao Lin}
\affiliation{International Center for Quantum Materials, School of Physics, Peking University, Beijing 100871, China}
\author{Ji Feng}
\email{jfeng11@pku.edu.cn}
\affiliation{International Center for Quantum Materials, School of Physics, Peking
University, Beijing 100871, China}
\affiliation{Hefei National Laboratory, Hefei 230088, China}
\date{\today}
\begin{abstract}
Our understanding of collective spin fluctuation in materials relies largely on Heisenberg-type spin Hamiltonians. Implicit in these spin models is the atomic moment picture that in transverse spin dynamics the magnetization around an atom undergoes precessional motion as a rigid moment, which has been challenged by emerging theoretical and experimental advances.
  To assess the validity of the atomic moment picture in spin dynamics, however, necessitates magnon wavefunctions from \textit{ab initio} methods without \textit{a priori} spin models.
  To this end, we develop an efficient model-free {\it ab initio} method for computing magnon spectrum and wavefunctions. Niu-Kleinman's adiabatic spin-wave dynamics is reformulated using linear perturbation theory into a generalized eigenvalue problem, which can be solved to produce magnon spectrum and wavefunctions without assuming atomic moments. We have implemented this method in the framework of density functional perturbation theory (DFPT). A dynamical extension of Niu-Kleinman equation of motion is proposed to improve inaccurate predicted magnon energies due to imperfect adiabaticity at higher energies. Based on so-obtained  {\it ab initio} magnon wavefunctions, we find the atomic moment picture to be valid in typical ferromagnets and antiferromagnets, but fails in the molecular orbital crystal Na$_2$IrO$_3$. Our results suggest that the usual spin Hamiltonian approach should be taken with a grain of salt, and possible experimental ramification on the issue is discussed. 
  

\end{abstract}

\pacs{}
\maketitle
\section{Introduction}
Magnons, or quanta of spin wave, are collective spin fluctuations and play important roles in the thermodynamics of magnetic materials and in the understanding of magnetism\cite{moriya2012spin}. Our
current understanding of magnons largely relies on the Heisenberg-type spin Hamiltonian
\begin{equation}
  \mathcal{H}=\sum_{mn} \vec{e}_{m}\cdot\mathcal{J}_{mn} \cdot\vec{e}_{n}.
\end{equation} 
where $\mathcal{J}_{mn}$ is a 2$^{nd}$ order tensor, and $\mathbf{e}_m$, $\mathbf{e}_n$ are unit vectors describing the direction of the magnetic moments of atoms at sites $m$ and $n$, respectively. The central assumption is that a magnetic crystal can be partitioned into cells surrounding magnetic atoms, within each of which the magnetization density is approximately unidirectional and can be treated as a rigid local moment along $\mathbf{e}_m$\cite{szilva2023quantitative}. In this atomic moment picture, the tensor $\mathcal{J}_{mn}$ encapsulates interaction between atomic moments, incorporating Heisenberg exchange interaction, single-ion anisotropy and Dzyaloshinskii-Moriya (DM) interaction, and so on. Magnetic phenomena observed in experiments can usually be accounted for by devising an amply sophisticated spin Hamiltonian. The interaction parameters, $\mathcal{J}_{mn}$, can also be computed using first principles methods, such as magnetic force theorm\cite{liechtenstein1984exchange,liechtenstein1985local,liechtenstein1987local}, cluster expansion\cite{drautz2004spin,singer2011spin,xiang2013magnetic} and spin-spiral calculation\cite{kubler1988density,sandratskii1998noncollinear}. These methods are all based on a mapping between first principles electronic energies and the classical energies of the spin Hamiltonian Eq.(1) for various atomic spin configurations. 

Traditionally, the atomic moment picture is adopted to describe magnetic materials with localized electronic states\cite{heisenberg1928}, but it seems to also do well with itinerant magnets\cite{turek2006exchange,szilva2023quantitative}. The ground state magnetization density predicted by {\it ab initio} calculations is indeed only significant near the nuclei of magnetic atoms for many magnets, localized or itinerant, seemingly justifying  the atomic moment pictures. The atomic moment picture suggests that the spins around a magnetic atom tend to remain parallel even in transverse motion ordered by the Hund's exchange coupling. However, emerging theoretical and experimental advances suggest interatomic electronic hopping can act in a way that compete with the Hund's exchange coupling. Mazin et al \cite{mazin20122,foyevtsova2013ab} argued that iridate Na$_2$IrO$_3$ is nearly itinerant and its $t_{2g}$ electron tends to be localized on an Ir$_6$ hexagon due to directional oxygen-mediated hoppings, which causes the formation of quasi-molecular orbitals fully delocalized over six Ir sites. Jin et al \cite{jin2023magnetic} found that in the itinerant magnet MnSi, the fundamental magnetic units involved  in incoherent magnetic scattering are best described by magnetic molecular orbitals sitting on an equilateral triangle of three Mn atoms.
The formation of quasi-molecular orbitals indicates specific linear combinations of orbitals across a cluster of atoms are tightly locked by interatomic hoppings. Orbitals on one atom can belong to quasi-molecular orbitals with different centers. When hopping outplays the Hund's rule coupling, it is unlikely for all magnetic orbitals on one atom to undergo coherent precessional motion as a single rigid spin, as assumed in the atomic moment picture.

Evidently, then, it is essential to assess the validity of the atomic moment picture. Nonetheless, a direct test of the possible breakdown of the atomic moment picture asks for an {\it ab initio} investigation of magnon wavefunctions in real materials, which is free of the atomic moment ansatz.  Another route to magnons is directly 
calculating the dynamical spin susceptibility, which is connected to the spin fluctuation spectrum by the fluctuation-dissipation theorem.
Magnon states appear as resonance peaks at magnon energies in the spin fluctuation spectrum.
Such calculations do not involve the atomic moment ansatz and can be accomplished using either time-dependent (td) DFPT \cite{savrasov1998linear,buczek2009energies,rousseau2012efficient,cao2018ab,gorni2018spin,tancogne2020time,singh2020ab,skovhus2021dynamic,liu2023implementation} or many-body perturbation theory \cite{karlsson2000spin,Ersoy2010,Okumura2019,Okumura2020}. However, the magnon wavefunction is not as readily available as in the spin Hamiltonian approaches. 


Earlier, Niu and Kleinman proposed an adiabatic theory of spin wave dynamics\cite{niu1998spin,niu1999adiabatic} that is suited for our purposes. In the adiabatic limit, many-electron wavefunction quenches instantaneously to the slowly evolving spin-wave configuration, or the "frozen spin-wave state", which is the lowest energy state that produce the required magnetization density. The spin wave dynamics is then formulated as the semiclassical evolution of the frozen spin-wave states. Niu and Kleinman's theory does not assume atomic moments, which in principle is suitable for investigating magnon wavefunctions in both localized and itinerant magnets. However, an {\it ab initio} solution of their adiabatic equations of motion (EOM) is hindered by a lack of the knowledge of spin-wave configurations. Consequently, the atomic moment ansatz is resorted to for construction of spin-wave states\cite{gebauer2000magnons,bylander2000fe}.

In this work, we develop a scheme that solves the Niu-Kleinman adiabatic EOM in the framework of linear perturbation theory to systematically obtain the magnon energies and wavefunctions. The Niu-Kleinman adiabatic EOM is reformulated as low energy solutions to an infinite-dimensional generalized eigensystem, which is then solved iteratively without resorting to the atomic moment ansatz. It is found  that the adiabatic theory overestimates the energies for higher energy magnons owing to imperfect adiabaticity. To overcome this problem, we propose a finite-frequency extension of the Niu-Kleinman adiabatic EOM to correct magnon energies. We have implemented this scheme at the level of td density-functional theory (DFT), which employs the first-order magnetization densities and wavefunctions calculated using DFPT techniques.

As demonstrations of our method, we conduct four case studies covering ferromagnetic (FM) and antiferromagnetic (AFM) materials, with and without significant Landau damping. We show that our method can accurately reproduce the magnon spectrum predicted by DFPT calculated spin fluctuation spectrum in systems without Landau damping. The magnon wavefunctions calculated for some typical magnets indeed conform to the atomic moment picture, but it is not the case for the molecular orbital crystal Na$_2$IrO$_3$. The spin-wave dynamics of the optical branch of magnon in Na$_2$IrO$_3$ cannot be described as the precession of rigid atomic moments, but instead involves significant inter-orbital transition. The breakdown of the atomic moment picture in this exemplifying underscores the interplay of Hund's exchange coupling and interatomic hopping, and suggests that caution is necessary when constructing effective spin models for multi-orbital systems. Our calculation also suggests spectroscopic signatures of such breakdown, or the lack of it.

This paper is organized as follows. As {\it ab initio} magnon wavefunctions are essential to obtaining a microscopic insights into the magnon dynamics, in order to assess the validity of the atomic moment picture, we will spend the first half of the paper describing our method to compute them without adopting a particular model. The method will then be applied to a few typical and not-so-typical magnets, to examine how close their spin dynamics is to the atomic moment picture. In Sec.\ref{sec:background}, we review the Niu and Kleinman's theory of adiabatic spin-wave dynamics and reformulate the adiabatic EOM using the generic linear perturbation theory of an interacting system. The meaning of adiabaticity for magnons is discussed, leading to a finite-frequency extension of adiabatic EOM that significantly improves the computed magnon energies for higher energy branches. In Sec.\ref{sec:implementation}, the implementation the adiabatic EOM in the DFPT framework, including computation of various matrix quantities, details of iterative eigensolver and the process of eigenvalue refinement. In Sec.\ref{sec:result}, we present results of magnon spectrum and wavefunction for FM half-metal CrO$_2$, AFM insulator NiO, body-centered-cubic(bcc) Fe and iridate Na$_2$IrO$_3$ without spin-orbit coupling (SOC), with an eye to whether and how they can be understood in the atomic moment picture. Finally, in \ref{sec:conclusion} our findings are summarized in light of possible new experimental and theoretical developments.

\section{Theoretical background\label{sec:background}}
\subsection{Niu and Kleinman's theory}
We start with a recapitulation of Niu and Kleinman's adiabatic spin wave theory\cite{niu1998spin,niu1999adiabatic}. Spin waves usually appear at much lower energies than other magnetic excitations (e.g. spin-flip electron-hole modes, aka the Stoner excitation); and thus the latter degrees of freedom (DOF) are assumed to be frozen out during the slow evolution of spin-waves. The electronic wavefunction quenches instantaneously to the spin-wave configuration, effectively being the lowest energy state that produce the magnetization density of the current spin-wave configuration. Such a "frozen spin-wave state" can be interpreted as an electronic ground state given a magnetization density in the context of spin density functional theory. Then the magnetization density is treated as slow semiclassical variables, whose EOM's are  derived from the adiabatic evolution of electronic wavefunction within the space of frozen spin-wave states.

Suppose we have a symmetry-broken ground state $|\Psi_0[\mathbf{m}_0]\rangle$ with magnetization density $\mathbf{m}_0(\mathbf{r})$. The excitation of spin waves leads to an instantaneous spin-wave configuration $\mathbf{m}(\mathbf{r})=\mathbf{m}_0(\mathbf{r})+\delta\mathbf{m}(\mathbf{r})$, with the corresponding frozen spin-wave states as a functional of $\boldsymbol m(\boldsymbol{r})$ denoted $|\Psi_0[\mathbf{m}]\rangle$. The adiabatic evolution of frozen spin-wave states can be formulated using the td variational principle, which leads to the following semiclassical EOM for $\mathbf{m}(\mathbf{r})$:
\begin{equation}
  \hbar \sum _{j}\int d\mathbf{r} '\Omega _{ij}(\mathbf{r} ,\mathbf{r} ')\frac{\partial }{\partial t} \delta m_{j}(\mathbf{r} 't) =\frac{\partial E[\mathbf{m}]}{\partial m_{i}(\mathbf{r} t)},\label{adiabatic}
\end{equation}
where 
\begin{equation}
  \Omega _{ij}(\mathbf{r} ,\mathbf{r} ')=\mathrm{i}\Big\langle \frac{\partial\Psi_0[\mathbf{m}]}{\partial m_i(\mathbf{r})}\Big|\frac{\partial\Psi_0[\mathbf{m}]}{\partial m_j(\mathbf{r}')}\Big\rangle-(i\leftrightarrow j)
  \label{eq:Omega}
\end{equation}
is the Berry curvature in the frozen spin-wave space, and $E[\mathbf{m}]=\langle\Psi_0[\mathbf{m}]|\hat{H}|\Psi_0[\mathbf{m}] \rangle$ is the energy of the spin wave configuration with $\hat{H}$ being the many-electron Hamiltonian. It should be emphasized that the formal definition Eq.\eqref{eq:Omega} is only meaningful when the derivative is carried out within the space of frozen spin-wave states. 

It was later pointed out by Qian and Vignale \cite{qian2002spin} that the adiabatic EOM Eq.\eqref{adiabatic} is a low frequency expansion truncated to linear order in $\omega$ of the following generic EOM 
\begin{equation}
  \sum _{j}\left[ \chi ^{-1}(\omega )\right]_{ij} \delta m_{j}=0, \label{exact}
\end{equation}
satisfied by spin fluctuation $\delta \mathbf{m}(\mathbf{r})$ due to the excited state at energy $\omega$. Here, $\chi(\omega)$ is the dynamical spin susceptibility, 
\begin{equation}
  \chi_{ij}(\mathbf{r},\mathbf{r}';\omega)=\frac{\mathrm{i}}{\hbar}\int_0^{\infty} dt e^{\mathrm{i}\omega t} \langle[\hat{m}_i(\mathbf{r}t),\hat{m}_j(\mathbf{r}')]\rangle , \label{chi} 
\end{equation}
($\hat{\mathbf{m}}(\mathbf{r})=\hat{\rho}(\mathbf{r})\mathbf{\sigma}$ is the magnetization density operator), that determines the magnetization density response $\delta \mathbf{m}(\omega)$ induced by external Zeeman field $\mathbf{B}(\omega)$, $\delta \mathbf{m}(\omega)=\chi(\omega)\mathbf{B}(\omega)$. Encoded in Eq.\eqref{exact} is the fact that an excited state can be driven by an infinitesimal perturbation at resonance. The connection between the adiabatic EOM Eq.\eqref{adiabatic} and generic EOM Eq.\eqref{exact} is established by the following relation:
\begin{equation}
  \mathrm{i}\Omega _{ij}(\mathbf{r} ,\mathbf{r} ') =\frac{\partial \left[ \chi ^{-1}\right]_{ij}(\mathbf{r} ,\mathbf{r} ';\omega )}{\partial \omega }\Bigl|_{\omega =0},
\end{equation}
and the stiffness theorem:
\begin{equation}
  \frac{\partial^2 E[\mathbf{m}]}{\partial m_i(\mathbf{r})\partial m_j(\mathbf{r}')}\Big|_{\delta \mathbf{m}=0}=[\chi^{-1}]_{ij}(\mathbf{r},\mathbf{r}';\omega=0).
\end{equation}
And Eq.\eqref{exact} truncated to the linear order in $\omega$
\begin{equation}
  \sum_{j}[\chi^{-1}]_{ij}\delta m_j=-\omega\sum_{j}\Big[
    \frac{\partial }{\partial \omega}
    \chi^{-1}\Bigl|_{\omega =0}\Big]_{ij}\delta m_{j},
\label{truncated1}
\end{equation}
with the notation  $\chi\equiv \chi(\omega=0)$, is Eq.\eqref{adiabatic} in the frequency domain.

Niu and Kleinman's adiabatic spin-wave theory does not involve the atomic moment ansatz, and is in this sense a better first principles approach than energy mapping methods. However, solving the adiabatic EOM Eq.\eqref{adiabatic} faces several challenges. The most essential one is a lack of the knowledge of the proper spin-wave configurations and the corresponding frozen spin-wave states. Previous efforts have been forced to invoke the atomic moment ansatz to build trial spin-wave configurations\cite{gebauer2000magnons,bylander2000fe}. In this case the adiabatic EOM formally reduces to the Landau-Lifshitz equations. Another practical difficulty lies in the gauge-invariant evaluation of magnon-space Berry curvature  Eq.\eqref{eq:Omega}, which has been accomplished by a finite-step Wilson loop calculation. In the following, we show how these difficulties can be naturally overcome by the adiabatic EOM reformulated in the linear perturbation theory.  

\subsection{Adiabatic equation of motion in linear perturbation theory}
With the operator derivative $\partial_{\omega}\chi^{-1}=-\chi^{-1}[\partial_{\omega}\chi]\chi^{-1}$, the adiabatic EOM Eq.\eqref{truncated1} can be recast into the following EOM of field $\mathbf{B}=\chi^{-1}\delta \mathbf{m}$:
\begin{equation}
  \sum_{j}\chi_{ij} B_j=\omega\sum_j\Big[\frac{\partial}{\partial\omega}\chi\Bigl|_{\omega =0}\Big]_{ij}B_j
  \label{truncated2}.
\end{equation}
In the following, we will derive a computable form of Eq.\eqref{truncated2} using the linear perturbation theory.

Consider the linear response of interacting electrons in a magnetic  crystal under perturbation $\delta H_{\mathrm{ex}}=-\mathbf{\sigma}\cdot\mathbf{B}$ due to a td external field of the Bloch-wave type:
\begin{equation}
  \mathbf{B}(\mathbf{r})=\mathbf{B}_{\mathbf{q}}(\mathbf{r})e^{\mathrm{i}(\mathbf{q}\cdot\mathbf{r}-\omega t)+\eta t}+\mathrm{c.c},
\end{equation}
where $\mathbf{q}$ is the crystalline momentum, $\mathbf{B}_{\mathbf{q}}(\mathbf{r})$ is lattice-periodic and $\eta\rightarrow 0^+$ is the switch-on parameter. The perturbed many-body electronic wavefunction to linear order is formally
\begin{equation}
  |\Psi\rangle = |\Psi _{0}[\boldsymbol{m}_0] \rangle +e^{-i\omega t}|\delta \Psi _{\mathbf{q}}(\omega) \rangle +e^{i\omega t}|\delta \Psi _{-\mathbf{q}}(-\omega) \rangle, 
\end{equation} 
where the first order wavefunctions $|\delta \Psi _{\mathbf{q}}(\omega) \rangle$ and $|\delta \Psi _{\mathbf{-q}}(-\omega) \rangle$ arise from
$\mathbf{B}_{\mathbf{q}}$ and $\mathbf{B}_{\mathbf{q}}^*$, respectively. The magnetization response will also be of Bloch wave type
\begin{equation}
  \delta \mathbf{m}(\mathbf{r}) =\delta \mathbf{m}_{\mathbf{q}}(\mathbf{r}\omega) e^{\mathrm{i}(\mathbf{q\cdot r}-\omega t)} +\mathrm{c.c.}
\end{equation}

We choose $\mathbf{B}_{\mathbf{q}}$ from the linear space spanned by a prescribed set of external fields $\{\mathbf{B}^a_{\mathbf{q}}\}$,
and denote the matrix representation of a generic tensor $\mathcal{K}$ in this space as 
\begin{equation}
  \begin{aligned}
    (\mathbf{B}^a_{\mathbf{q}}|&\mathcal{K}|\mathbf{B}^b_{\mathbf{q}}) \\
    &\equiv \sum_{ij} \iint d\mathbf{r} d\mathbf{r}' B^{a*}_{\mathbf{q},i}(\mathbf{r}) \mathcal{K}_{ij}(\mathbf{r},\mathbf{r}') B^b_{\mathbf{q},j}(\mathbf{r}') e^{-i\mathbf{q}\cdot(\mathbf{r}-\mathbf{r}')}.   
  \end{aligned}
\end{equation}
One can show the matrix representations of $\chi$ and $\frac{1}{\hbar}{\partial_{\omega}}\chi$,
\[\begin{aligned}
  S_{ab}(\omega) &\equiv (\mathbf{B}^a_{\mathbf{q}}|\chi(\omega)|\mathbf{B}^b_{\mathbf{q}}),  \\
  H_{ab}(\omega) &\equiv (\mathbf{B}^a_{\mathbf{q}}|\frac{1}{\hbar}\frac{\partial}{\partial\omega}\chi(\omega)|\mathbf{B}^b_{\mathbf{q}}),
\end{aligned}\]
(almost) satisfy the following equations:
\begin{subequations}
\begin{align}
  S_{ab}(\omega)&=\int d\mathbf{r} \mathbf{B}^{a*}_{\mathbf{q}}(\mathbf{r})\cdot\delta\mathbf{m}^{b}_{\mathbf{q}}(\mathbf{r}\omega), \label{Sab} \\
  H_{ab}(\omega)&=\langle\delta\Psi^a_{\mathbf{q}}(\omega)|\delta\Psi^b_{\mathbf{q}}(\omega)\rangle-\langle\delta\Psi^b_{-\mathbf{q}}(-\omega)|\delta\Psi^a_{-\mathbf{q}}(-\omega)\rangle,\label{Hab}
\end{align}
\end{subequations}
despite some subtlety for resonant excitations (see Appendix \ref{AppA} for a more rigorous treatment).  
The matrix representation of the adiabatic EOM Eq.\eqref{truncated2} then becomes the following generalized eigensystem:
\begin{equation}
  \sum_b\hbar\omega H_{ab}\lambda_b=\sum_b S_{ab}\lambda_{b},\label{generalized},
\end{equation} 
where $H\equiv H(0)$ and $S\equiv S(0)$ is understood. A more straightforward derivation of Eq.\eqref{generalized} can be made using the td variational principle as in Niu-Kleinman's formulation by considering adiabatic evolution of frozen spin-wave states in the linear perturbated regime around the ground state.

If the space $\{\delta\mathbf{m}^a_{\mathbf{q}}\}$ conincides with the tangent space of spin-wave configurations around the ground state magnetization $\mathbf{m}_0$, Eq.\eqref{generalized} describes the Niu-Kleinman adiabatic spin-wave dynamics and the eigenvalues $\hbar\omega=\pm\hbar\omega_{\pm\mathbf{q}}$ yields the energy $\hbar\omega_{\mathbf{q}}$ ($\hbar\omega_{-\mathbf{q}}$) of every $\mathbf{q}$ ($-\mathbf{q}$) magnon exited by $\mathbf{B}_{\mathbf{q}}$ ($\mathbf{B}^*_{\mathbf{q}}$). An added advantage is that we need not worry about gauge freedom in the evaluation of Berry curvature Eq.\eqref{eq:Omega}, since the gauge of first-order wavefunctions $|\delta\Psi^{a}_{\pm\mathbf{q}}\rangle$'s is already fixed by the shared ground state $|\Psi_{0}\rangle$. 

We face the same problem of lacking the knowledge of proper spin-wave configurations. The situation is different from the case of phonons, another commonplace bosonic excitations, their DOF expand a known finite dimensional linear space of ionic displacements at each crystal momentum $\mathbf{q}$. In the magnon case, $\delta\boldsymbol{m}$ corresponds to slow, continuous electronic DOF, for which we only have little intuitions beyond the atomic moment picture.
This problem can be overcome by solving the EOM Eq.\eqref{generalized} (or its operator form Eq.\eqref{truncated2}) in the infinite-dimensional space $\{\delta\mathbf{m}^a_{\mathbf{q}}\}$ of arbitrary magnetization response. We argue that proper spin-wave configurations will correspond to low energy solutions of this infinite-dimensional eigensystem. Since the EOM Eq.\eqref{truncated2} is a low-frequency expansion of the exact EOM Eq.\eqref{exact}, low energy solutions to Eq.\eqref{truncated2} should adequately approximate those to Eq.\eqref{exact}, which are expected to be the magnon modes. Consequently, we can obtain {\it ab initio} spin-wave configurations by solving Eq.\eqref{generalized} without resorting to the atomic moment ansatz. 

\subsection{Adiabaticity and dynamical extension}

Our derivation of the adiabatic EOM Eq.\eqref{generalized} in the linear perturbation theory also reveals the essential requirement of adiabaticity for magnons and allows a dynamical extension of the EOM at finite frequencies, as we now clarify.

First, let us discuss the meaning of adiabaticity for magnons, which is slightly different with the adiabaticity for phonons. Within the Born-Oppenheimer approximation, the DOF of phonons are non-electronic and the adiabaticity of phonons means electronic DOF are not excited as ions move. The DOF of magnons $\delta\mathbf{m}(\mathbf{r})$, however, are themselves electronic. The adiabaticity of magnons amounts to that adiabatic dynamics of magnetization really describes magnon dynamics, or to what extent a frozen spin-wave state $|\Psi_0[\mathbf{m}]\rangle$ looks like the ground state $|\Psi_0[\mathbf{m}_0]\rangle$ with only magnon excitations. If Stoner excitations are inevitably involved in $|\Psi_0[\mathbf{m}]\rangle$, the dynamics of magnon alone is not completely adiabatic. To quantify this adiabaticity, we define state-resolved spin fluctuation 
\[\delta \mathbf{m}_{n\pm\mathbf{q}}(\mathbf{r})=\langle \Psi_0|\hat{\mathbf{m}}(\mathbf{r})|\Psi_{n\pm\mathbf{ q}}\rangle\langle\Psi_{n\pm\mathbf{ q}}|\delta\Psi_{\pm\mathbf{q}}\rangle,\]
where $|\Psi_{n\pm\mathbf{ q}}\rangle$ is the $n$th excited state with crystal momentum $\pm\mathbf{q}$, then Eq.\eqref{truncated2} can be rewritten as 
\begin{equation}
\begin{aligned}
\frac{1}{\hbar\omega}\Big[\sum_n &\delta\mathbf{m}_{n\mathbf{q}}+\sum_m\delta\mathbf{m}^*_{m\mathbf{-q}})\Big]=\\
&\sum_n \frac{1}{E_{n\mathbf{q}}-E_0}\delta\mathbf{m}_{n\mathbf{q}}+\sum_m\frac{1}{E_0-E_{m-\mathbf{q}}}\delta\mathbf{m}^*_{m\mathbf{-q}}.    
\end{aligned}\label{harmonic}
\end{equation}
Thus the eigenvalue $\hbar\omega$ is seen to be a harmonic mean of excitation energies $\{E_{n\mathbf{q}}-E_0\}$ and $\{E_0-E_{m-\mathbf{q}}\}$ weighted by spin fluctuation $\{\delta\mathbf{m}_{n\mathbf{q}}\}$ and $\{\delta\mathbf{m}_{m-\mathbf{q}}^*\}$. The adiabatic spin wave theory predicts $\hbar\omega=E_{n\mathbf{q}}-E_0$ or $E_0-E_{m-\mathbf{q}}$ with $|\Psi_{n\mathbf{q}}\rangle$ or $|\Psi_{m\mathbf{-q}}\rangle$ being the electronic state with a single excited magnon. But this can only be achieved when spin fluctuation from other excited states is negligible compared to $\delta\mathbf{m}_{n\mathbf{q}}$ or $\delta\mathbf{m}_{m-\mathbf{q}}^*$. 
The more dominant contribution a magnon can make to the magnetization response, the more accurate is
the energy predicted by the adiabatic theory. Thus, perfect adiabaticity translates to the absence of Stoner excitations. Even though the Stoner excitations may be inevitable under perturbation, a magnon with much lower energy than Stoner excitations can be nearly adiabatic, as $\delta\mathbf{m}_{n\pm\mathbf{q}}$ implicitly contains a $\frac{1}{E_0-E_{n\pm\mathbf{q}}}$ factor. But when the magnon energy is closer to Stoner excitations, predictions of the adiabatic theory can be inaccurate.


Second, Eq.\eqref{truncated2} can be readily generalized to finite frequencies,
\begin{equation}
\sum_j\chi_{ij}(\omega_0)B_j = (\omega-\omega_0)\sum_j\Big[\frac{\partial}{\partial \omega}\chi(\omega)\Big|_{\omega=\omega_0}\Big]_{ij}B_j,\label{truncated3}
\end{equation}
with a linear expansion of Eq.\eqref{exact} at frequency $\omega_0$ assuming $\chi^{-1}(\omega)$ analytic near resonance. The matrix representation of Eq.\eqref{truncated3} can be readily calculated according to Eq.\eqref{Sab} and Eq.\eqref{Hab} as in the static case. This leads to a dynamical extension of the adiabatic EOM Eq.\eqref{generalized} at finite frequencies. Eq.\eqref{truncated3} provides a more accurate description of higher energy magnons if the chosen $\omega_0$ is close to resonance (as contributions of these magnons to the magnetization response are enhanced). It leads to a formula for energy correction [see Eq.\eqref{correction} below], which allows an accurate reproduction of DFPT calculated magnon spectrum, as will be shown in Sec.\ref{sec:result}.

\section{Implementation \label{sec:implementation}}
In this section, we discuss how  the adiabatic EOM Eq. \eqref{generalized} is implemented in DFPT within the adiabatic td DFT. We then describe the workflow of solving for {\it ab initio} magnon spectrum and wavefunctions, providing essential technical details of the iterative eigensolver of the adiabatic EOM and the eigenvalue refinement based on Eq.\eqref{truncated3}. 

\subsection{Implementation in DFPT framework}
Kohn-Sham DFT can be viewed as a mean field theory of interacting electrons that produces exact ground state density\cite{kohn1965self}. The td Kohn-Sham theory within the adiabatic approximation\cite{zangwill1980density,zangwill1980resonant} describes the electronic dynamics under the instantaneous mean fields. The perturbation theory within the td Kohn-Sham DFT, i.e. td DFPT, allows the calculation of dynamical density response (including spin susceptibility $\chi$ or its matrix representation $S_{ab}$) of many-electron systems in real materials. Below, we show that $H_{ab}$ in Eq.\eqref{generalized} can be conveniently computed in the same framework.

Perturbation theory in first order is solved through the following Sternheimer equation\cite{sternheimer1954electronic,savrasov1998linear,baroni2001phonons}:
\begin{equation}
  (\mathrm{i}\partial_t-\hat{H})|\delta\Phi(t)\rangle=\delta \hat{H}(t)|\Phi_{0}(t)\rangle\label{stein}
\end{equation}
where $|\Phi_{0}(t)\rangle=e^{-\mathrm{i}E_0 t}|\Phi_{0}\rangle$ is the ground state of the Kohn-Sham Hamiltonian $\hat{H}$ with energy $E_0$ and $|\delta\Phi(t)\rangle$ stands for the first order wavefunction. The perturbation $\delta H(t)$ felt by electrons includes both the external field $v_{\text{ext}}(t)$ and a modification of the mean field $v_{\text{ind}}(t)$ due to (4-component ) density response $\delta\rho(t)$:
\begin{equation}
  v_{\text{ind}}(\mathbf{r}t)=\int d\mathbf{r}' f(\mathbf{r},\mathbf{r}')\delta\rho(\mathbf{r}'t).
\end{equation}
where $f$ is the interaction kernel. This leads to the well-known Dyson equation that relates the exact density response function $\chi(\omega)$ to the bare density response $\chi_0(\omega)$ of the Kohn-Sham system:

\begin{equation}
  \chi(\omega)^{-1}=\chi_0(\omega)^{-1}-f. \label{RPA}
\end{equation}
Since the interaction kernel $f$ bears no time-dependence under the adiabatic approximation, we have
\begin{widetext}
\begin{equation}
  \mathrm{i}\Omega _{ij}(\mathbf{r} ,\mathbf{r} ') =\frac{\partial \left[ \chi ^{-1}\right]_{ij}(\mathbf{r} ,\mathbf{r} ';\omega )}{\partial \omega }\Bigl|_{\omega =0}=\frac{\partial \left[ \chi_{0} ^{-1}\right]_{ij}(\mathbf{r} ,\mathbf{r} ';\omega )}{\partial \omega }\Bigl|_{\omega =0}=\mathrm{i}\left\langle \frac{\partial\Phi_{0}[\mathbf{m}]}{\partial m_i(\mathbf{r})}\Big|\frac{\partial\Phi_{0}[\mathbf{m}]}{\partial m_j(\mathbf{r})}\right\rangle-(i\leftrightarrow j), \label{Omega_dfpt}
\end{equation}
\end{widetext}
where $\Phi_{0}[\mathbf{m}]$ is an non-interacting ground state with the same $\mathbf{m}(\mathbf{r})$ as the frozen spin-wave state $\Psi_{0}[\mathbf{m}]$, i.e. the corresponding Kohn-Sham ground state. Eq.\eqref{Omega_dfpt} suggests that the interacting system and the corresponding Kohn-Sham system have the same magnon-space Berry curvature, with the adiabatic interaction kernel. Consequently, matrix representation $H_{ab}$ in Eq.\eqref{Hab} can be conveniently calculated by replacing $|\delta \Psi_{\pm\mathbf{q}}\rangle$ with their DFPT counterpart $|\delta \Phi_{\pm\mathbf{q}}\rangle$. A more careful proof of this point is provided in the Appendix.\ref{AppA}.

\begin{figure*}[ht]
  \includegraphics[width=180 mm]{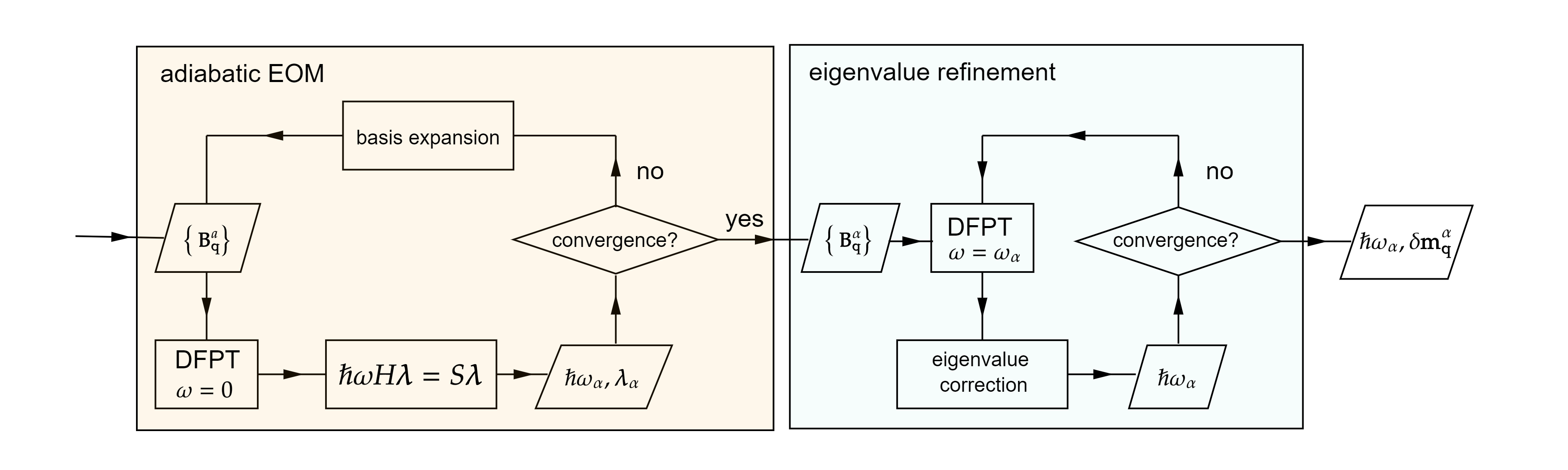}
  \caption{Work flow of our method. The precedure is comprised of two parts: solution of the adiabatic EOM and further eigenvalue refinement. See detailed descriptions in the main text.}
  \label{fig:flow}
\end{figure*}
\subsection{The workflow}

The two-step workflow to solve the {\it ab initio} magnon spectrum and wavefunction of a magnetic crystal is depicted in Fig.\ref{fig:flow}. In the first step, static $\omega=0$ DFPT calculations are performed to calculate the matrix representation of the generalized eigensystem Eq.\eqref{generalized} within a trial subspace $\{\mathbf{B}^{a}_{\mathbf{q}}\}$. The low energy eigenvalues and eigenvectors $\{\hbar\omega_{\alpha},\mathbf{\lambda_\alpha}\}$ of the eigensystem are iteratively solved aided by a basis expansion algorithm to be elaborated below. In the second step, dynamical DFPT calculation is performed at the current eigenfrequency $\omega_{\alpha}$ for corresponding 
$\mathbf{B}^{\alpha}_{\mathbf{q}}=\sum_a \lambda_{\alpha,a}\mathbf{B}^{a}_{\mathbf{q}} $ to compute a magnon energy correction according to Eq.\eqref{truncated3}. 
This refinement process can be iteratively performed on the updated eigenvalues, continously improving them until the convergence is reached. The final eigenvalues and magnetization response $\{\delta \mathbf{m}^{\alpha}_{\mathbf{q}}\}$ at the end of the refinement process form the output magnon spectrum and wavefunctions. This procedure is developed upon our previous implementation of spinor DFPT \cite{liu2023implementation} in the projector augmented-wave (PAW) pseudopotential framework based on Vienna Ab-initio Simulation Package (VASP)\cite{kresse1996efficient}, which solves the Sternheimer Equation Eq.\eqref{stein} under td electromagnetic fields to produce first-order wavefunction $|\delta\Phi\rangle$ as well as linear density response $\delta\rho$. In the following, we discuss the details of the basis-expansion algorithm and calculation of eigenvalue correction.

\subsubsection{Iterative solution of adiabatic EOM}
As mentioned in the previous section, we find the spin-wave configuration by solving the low-energy eigenmodes of the generalized eigensystem Eq.\eqref{generalized}. Typical iterative eigensolvers use basis expansion method and in each iteration\cite{wood1985new}
\begin{enumerate}
  \item solve the eigensystem in a subspace as the best approximation of the exact eigenvalues and eigenvectors,
  \item properly expand the subspace based on the approximated eigenvalues and eigenvectors.
\end{enumerate} 
The eigenvalues and eigenvectors are improved until convergence. The generalized eigensystem in subspace is Hermitian and can be solved by the conventional Choleski-Householder procedure. Solving non-Hermitian eigensystem $S^{-1}H$ is also feasible and particularly useful for calculation at $\Gamma$, where magnon usually appears as zero-energy Goldstone mode and cause diverging response. In this case, DFPT should be performed with a finite $\eta$, with which $\chi$ contains an significant anti-Hermitian component. A detailed discussion of solution of adiabatic EOM with finite $\eta$ is provided in Appendix \ref{AppA}.

Typical iterative eigensolvers expand the subspace based on preconditioned residual vectors $\mathcal{K}\mathbf{R}$. The residual vector in our case reads
\[\mathbf{R}^{\alpha}=\Big[\omega^{\alpha}\frac{\partial}{\partial\omega}\chi-\chi\Big]\mathbf{B}^{\alpha},\] for which a good preconditioning operator $\mathcal{K}$ should satisfy 
\[\mathcal{K}\approx \Big[\omega^{\alpha}\frac{\partial}{\partial\omega}\chi-\chi\Big]^{-1}.\] Unfortunately, we cannot calculate the operation of ${\partial_\omega}\chi$ on an external field $\mathbf{B}_{\text{app}}^{\alpha}$ but only its matrix representation $H_{ab}$ in Eq.\eqref{Hab} within a subspace. So we are not able to directly measure whether the solved eigenvector is really close to exact. Since we are only interested in low energy eigenvectors, $\mathcal{K}=\chi^{-1}$ looks to be a good preconditioning operator. Moreover, note the following identity exists
\[\Big[\chi^{-1}\frac{\partial}{\partial \omega}\chi\Big]\mathbf{B}=-\Big[\frac{\partial}{\partial \omega}\chi^{-1}_0\Big]\delta \mathbf{m}=\chi_0^{-1} \Big[\frac{\partial}{\partial \omega}\chi_0\Big](\mathbf{B}+f\delta \mathbf{m}),\]
where we have used the Dyson equation Eq.\eqref{RPA}. If the Stoner excitation has a well established gap $\Delta E$, then we can approximate ${\partial_\omega}\chi_0\sim \chi_0/\Delta E$, and the preconditioned residual vectors will be linear combinations of $\mathbf{B}^{\alpha}$ and $f\delta \mathbf{m}^{\alpha}$. In other words, the induced field $f\delta \mathbf{m}^{\alpha}$'s form a good set of vectors for basis expansion and are added to the subspace $\{\mathbf{B}^{a}_{\mathbf{q}}\}$ in each iteration.  

\subsubsection{Eigenvalue refinement}
In magnetic crystals without Landau damping, the solution of adiabatic EOM usually accurately reproduce the low energy part ($<\frac{1}{3}$ Stoner excitation gap, according to our tests) of the magnon spectrum predicted by DFPT dynamical spin susceptibility calculation. However, for the higher energy part of magnon spectrum, the magnon energy is usually slightly overestimated. This may be due to inaccurate eigenfields $\mathbf{B}^{\alpha}$'s as we cannot check their convergence by calculating their residual vector $\mathbf{R}^{\alpha}$'s. The adiabaticity of spin-wave dynamics is also not well justified when the magnon energy is smaller but comparable to that of low-lying Stoner excitations. In either case, the solved eigenfield $\mathbf{B}^{\alpha}$ inevitably induces a few Stoner excitations at higher energies, which causes the observed energy overestimation,[c.f. Eq.\eqref{harmonic}]. 

According to Eq.\eqref{harmonic}, the accuracy of predicted magnon energy depends on whether the magnon contribution to the magnetization response $\delta \mathbf{m}^{\alpha}$ is dominant. This is also true for the extended adiabatic EOM at finite frequency Eq.\eqref{truncated3}. The closer a $\omega_0$ is to a magnon energy, the more dominant the magnon contribution is in Eq.\eqref{harmonic}.
This leads to the following refinement process of eigenvalues. We perform the DFPT calculation for td external field, $\mathbf{B}_{\mathbf{q}}^{\alpha}e^{-i\omega_{\alpha}t}+\mathrm{c.c}$, at eigenfrequency $\omega_{\alpha}$ solved from the adiabatic EOM, 
to get magnetization response $\delta \mathbf{m}_{\mathbf{q}}^{\alpha}(\omega_{\alpha})$ and first order wavefunctions $|\delta \Phi^{\alpha}_{\pm\mathbf{q}}(\pm \omega_{\alpha})\rangle$. Then the corrected magnon energy is calculated, $\hbar\omega_{\alpha}\leftarrow\hbar\omega_{\alpha}+\hbar\delta \omega_{\alpha}$, with energy correction
\begin{equation}
  \hbar\delta\omega_{\alpha}= S_{\alpha\alpha}(\omega_{\alpha})/H_{\alpha\alpha}(\omega_{\alpha}),\label{correction}
\end{equation}
where 
\begin{subequations}
  \begin{align}
    H_{\alpha \alpha}(\omega)=&\left\langle\delta \Phi_{\boldsymbol{q}}^\alpha\left(\omega\right) \mid \delta \Phi_{\boldsymbol{q}}^\alpha\left(\omega\right)\right\rangle\notag\\
    &-\left\langle\delta \Phi_{-\boldsymbol{q}}^\alpha\left(-\omega\right) \mid \delta \Phi_{-\boldsymbol{q}}^\alpha\left(-\omega\right)\right\rangle\\
    S_{\alpha\alpha}(\omega) =&\int d\mathbf{r} \mathbf{B}_{\mathbf{q}}^{\alpha*}(\mathbf{r}) \cdot \delta \mathbf{m}_{\mathbf{q}}^{\alpha}(\mathbf{r}\omega).        
  \end{align}
\end{subequations}
This refinement process can be iterated on the updated eigenvalues until convergence is reached.

\section{Application to magnetic materials\label{sec:result}}

Our method is now applied to three collinear ferromagnets and an antiferromagnet, with or without significant Landau damping. In Sec.\ref{cro2} and \ref{nio}, we present the results for ferromagnet CrO$_2$ and antiferromagnet NiO. They are canonical examples where the Stoner excitations have a sizable gap and the adiabatic assumption holds. In Sec.\ref{bccFe}, we investigate performance of the adiabatic EOM in body-centered cubic (bcc) Fe, where low-lying Stoner excitations overlap with magnon in energy, and leads to a significant non-adiabatic effect.
In Sec.\ref{nairo}, we study the non-trivial case of ferromagnetic Na$_2$IrO$_3$, where our results enable a microscopic discussion of the fundamental magnetic units. Evidence of breakdown of the atomic moment picture is discussed in detail.

Ground state DFT and DFPT calculations are performed within local spin-density approximation \cite{perdew1981} using a plane-wave cutoff 500eV. SOC is not considered and the ground state magnetization of these materials is chosen to be along the $z$-axis. Other computational settings are summarized in TABLE.\ref{table:setting}. The initial set of external fields for solution of adiabatic EOM is chosen to be $\mathbf{B}_{\mathbf{q}}(\mathbf{r})=B(\mathbf{r})\hat{\mathbf{n}}$; for each magnetic atom, two $\mathbf{B}_{\mathbf{q}}$ are used with with $B(\mathbf{r})$ being an Gaussian centered at this atom and $\hat{\mathbf{n}}$ along $\hat{\mathbf{x}}$ and $\hat{\mathbf{y}}$ directions. We also carried out the DFPT calculations of spin fluctuation spectrum to provide the benchmark magnon spectra $\hbar\omega^{\chi}_{\mathbf{q}}$. In these calculations, the external field is chosen to be similar atomic Gaussian field, 
$\mathbf{B}_{\mathbf{q}}(\mathbf{r})=B(\mathbf{r})(\hat{\mathbf{x}}-\mathrm{i}\hat{\mathbf{y}})$ on one magnetic atom. The resultant magnetization response $\delta m_{\mathbf{q},+}(\omega)=\delta m_{\mathbf{q},x}(\omega)+\mathrm{i}\delta m_{\mathbf{q},y}(\omega)$ is intergated with $B(\mathbf{r})$ to produce 
\[\chi_{+-}(\mathbf{q}\omega)\equiv\int d\mathbf{r} B(\mathbf{r})\delta m_{\mathbf{q},+}(\mathbf{r}\omega),\]
whose imaginary part Im$\chi_{+-}(\mathbf{q}\omega)$ is the reported spin fluctuation spectrum. The benchmark magnon spectrum $\hbar\omega^{\chi}_{\mathbf{q}}$ is obtained from Im$\chi_{+-}(\mathbf{q}\omega)$ by multi-peak Lorentz fitting.

\begin{table*}[htpb]
  \centering
  \setlength{\tabcolsep}{8mm}{
  \begin{threeparttable}[b]
  \caption{Computational settings of DFT and DFPT calculation. $\mathbf{k}$-mesh is chosen to be $\Gamma$-centered. $\eta$ is the switch-on parameter for DFPT calculations. $\sigma$ is the width of the Gaussian smearing of Fermi-Dirac distribution.}
  \label{table:setting}
  \begin{tabular}{cccccc}
  \hline
  \hline
  Material & $\mathbf{k}$-mesh & $\eta$ (meV) & $\sigma$ (meV) & LDA+$U$\\
  \hline
  \makecell[l]{CrO$_2$} &$10\times10\times10$ & 10 & 50  & No\\
  \makecell[l]{NiO} &$10\times10\times10$ & 10 & 50  & Yes\\
  \makecell[l]{bcc Fe} &$30\times30\times30$ & 20 & 20  & No\\
  \makecell[l]{Na$_2$IrO$_3$} &$10\times10\times10$ & 10 & 50  & Yes\\
  \hline
  \hline
  \end{tabular}
  \end{threeparttable}
  }
  \end{table*}
\subsection{Ferromagnet CrO$_2$ \label{cro2}}
\begin{figure}
  \includegraphics[width=80 mm]{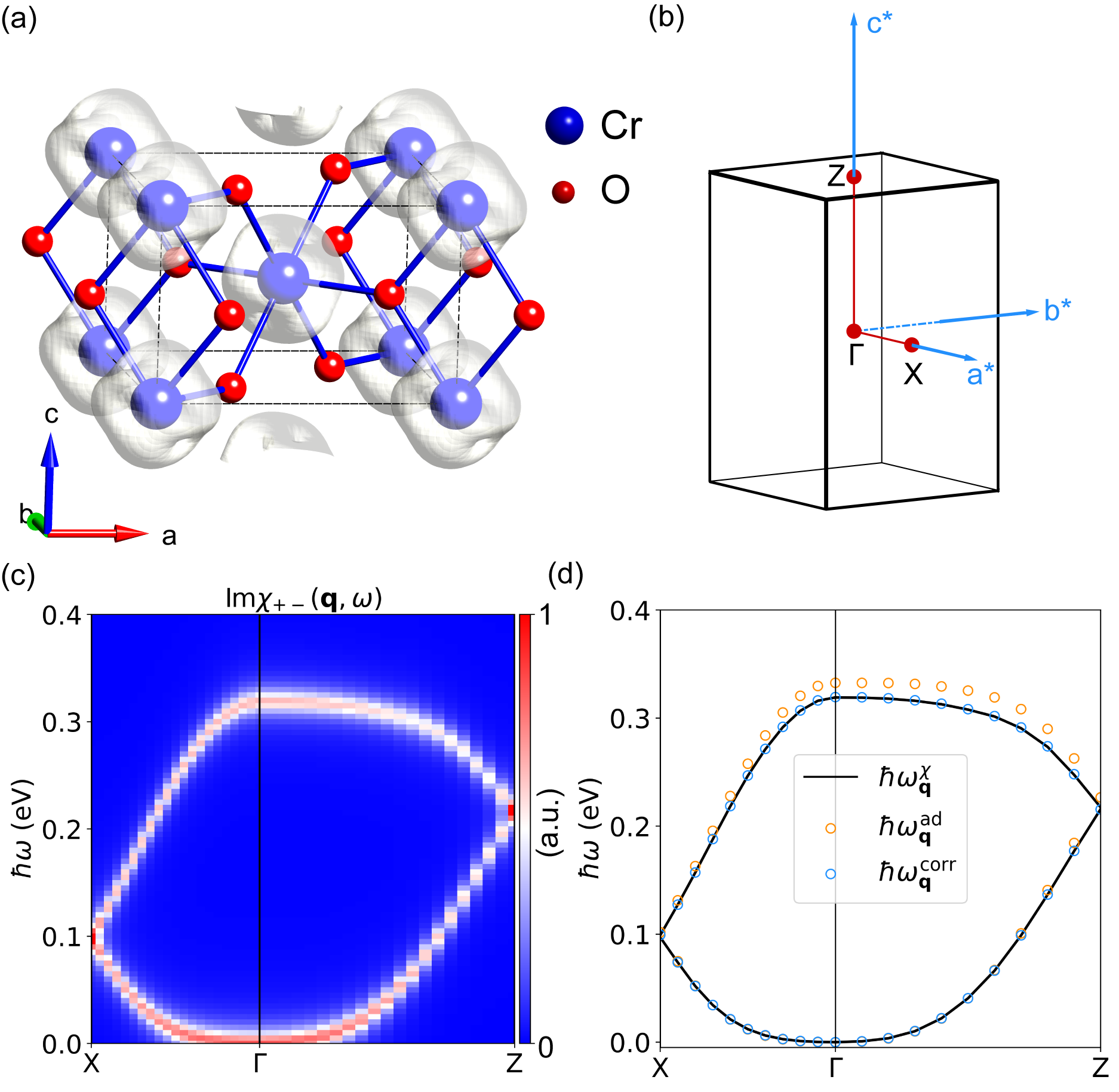}
  \caption{(a) The crystal structure of CrO$_2$ plotted with an isosurface of ground state $z$-magnetization. (b)The Brillouin zone with  high symmetry paths highlighted. (c) The spin fluctuation spectrum Im$\chi_{+-}(\mathbf{q}\omega)$. (d) Adiabatic spectrum $\hbar\omega^{\text{ad}}_{\mathbf{q}}$ and corrected spectrum $\hbar\omega^{\text{corr}}_{\mathbf{q}}$ plotted with fitted magnon spectrum $\hbar\omega^{\chi}_{\mathbf{q}}$}
  \label{fig:cro_chi}
\end{figure}

Chromium dioxide CrO$_2$ is a half-metallic ferromagnet with a rutile crystal structure. As shown in Fig.\ref{fig:cro_chi}(a), each Cr atom is situated at the center of an octahedral cage formed by oxygen atoms. 
The experimental lattice parameters $a = b = 4.422 $\AA~ and $c = 2.918$ \AA~ \cite{cloud1962x} are used in our calculations. 

The spin fluctuation spectrum Im$\chi_{+-}(\mathbf{q}\omega)$ of CrO$_2$ in Fig.\ref{fig:cro_chi}(c), shows two branches of magnons as a unit cell has two magnetic Cr atoms. The acoustic branch displays quadratic dispersion near $\Gamma$ as predicted by the linear spin-wave theory for ferromagnets. The Stoner excitations are found only above 400 meV along $\Gamma$-X/Z, and do not overlap with magnons in energy\cite{liu2023implementation}. Thus, the Landau damping is negligible and the adiabatic assumption is well satisfied. The adiabatic spectrum $\hbar\omega^{\text{ad}}_{\mathbf{q}}$ solved from Eq.\eqref{generalized} is plotted along with the benchmark spectrum $\hbar\omega^{\chi}_{\mathbf{q}}$ plotted in Fig.\ref{fig:cro_chi}(d). Despite general resemblance, $\hbar\omega^{\text{ad}}_{\mathbf{q}}$ overestimates magnon energies in the optical branch compared to $\hbar\omega^{\chi}_{\mathbf{q}}$, by as much as 10 meV. Upon the correction by Eq.\eqref{correction}, the corrected spectrum $\hbar\omega^{\text{corr}}_{\mathbf{q}}$ agrees perfectly with $\hbar\omega^{\chi}_{\mathbf{q}}$.

\begin{figure}
  \includegraphics[width=80 mm]{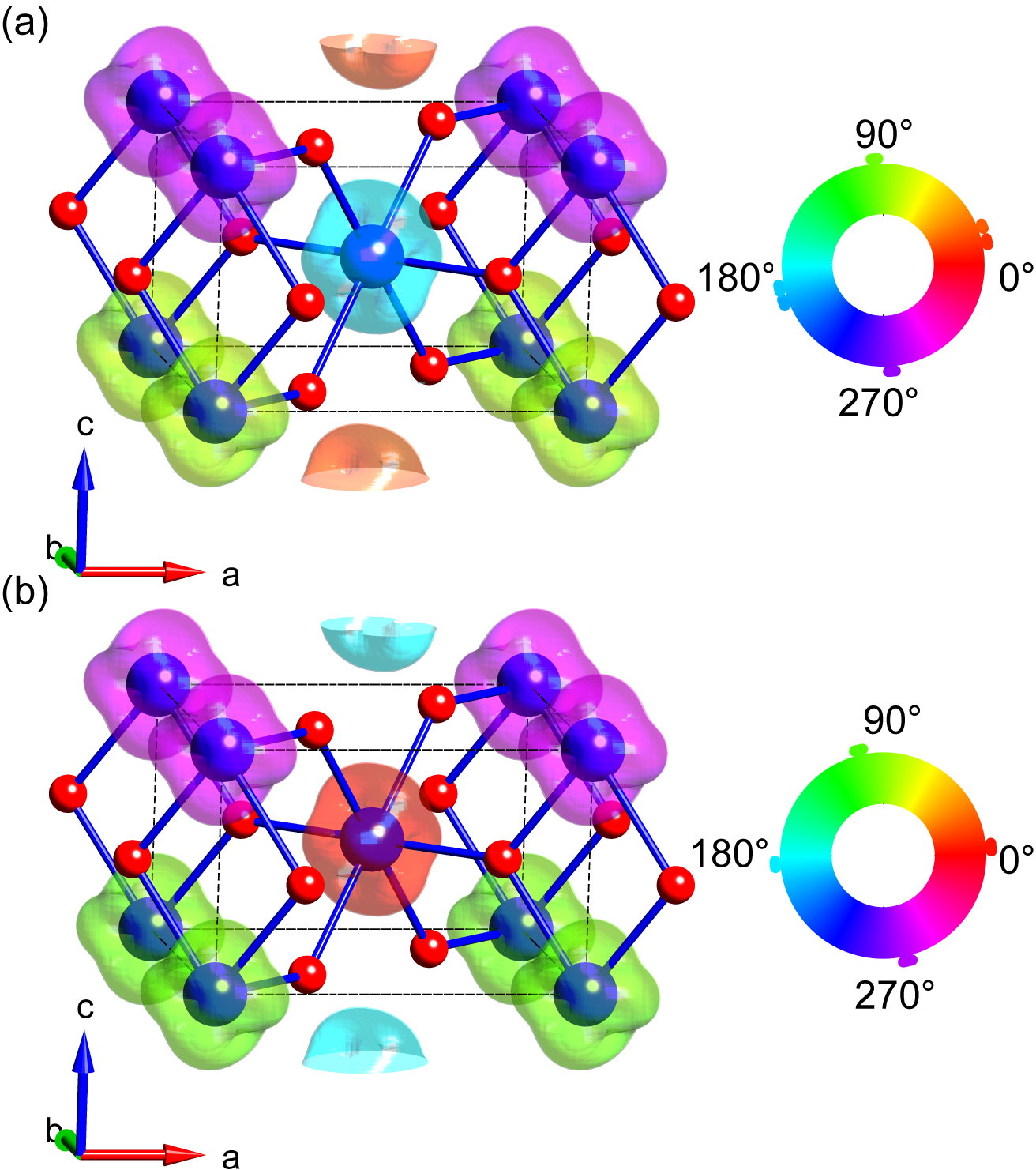}
  \caption{Magnon wavefunctions of CrO$_2$ at Z. (a) and (b) display isosurface of $|\delta \mathbf{m}_{\mathbf{q}}(\mathbf{r})e^{\mathrm{i}\mathbf{q}\cdot\mathbf{r}}|$ of two magnon modes, respectively. The isosurfaces are colored according to direction of Re$[\delta \mathbf{m}_{\mathbf{q}}(\mathbf{r})e^{\mathrm{i}\mathbf{q}\cdot\mathbf{r}}]$ within $x-y$ planes, whose polar angles are marked on the color wheels beside. }
  \label{fig:cro_wave}
\end{figure}

Fig.\ref{fig:cro_wave} (a) and (b) show the two magnon wavefunctions $\delta \mathbf{m}_{\mathbf{q}}(\mathbf{r})e^{\mathrm{i}\mathbf{q}\cdot\mathbf{r}}$ of CrO$_2$ at high symmetry point Z ($\frac{1}{2}\mathbf{c}^*$). The spin fluctuation on two Cr atoms related by translation $\mathbf{c}$ are anti-phase as expected. The spatial distribution of the magnon wavefunctions characterized by non-spherical clouds localized around each Cr atom coincide with that of the ground state magnetization shown in Fig.\ref{fig:cro_chi}(a). Moreover, the magnetization in these magnons around each Cr atom is almost unidirectional. All these features conform to the atomic moment picture, where magnetization around each Cr atom undergoes precessional motion as a rigid moment in a spin-wave mode.

\subsection{Antiferromagnet NiO\label{nio}}
\begin{figure}
  \includegraphics[width=80 mm]{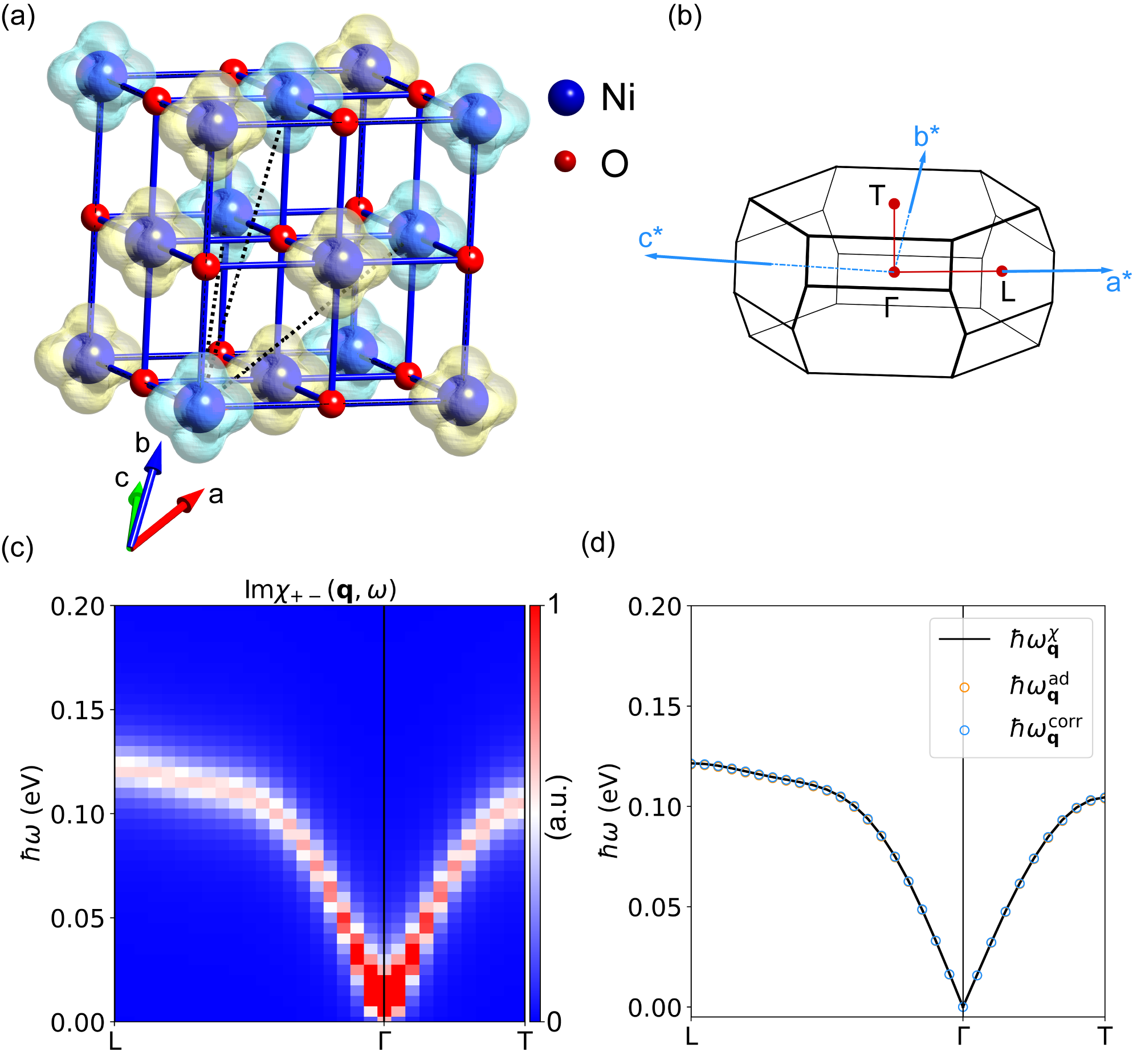}
  \caption{(a) The crystal structure of NiO plotted with an isosurface of ground state $z-$magnetization. Cyan/yellow indicates spin-up/down, respectively. (b) The Brillouin zone with high symmetry paths highlighted. (c) The spin fluctuation spectrum Im$\chi_{+-}(\mathbf{q}\omega)$. (d) Adiabatic spectrum $\hbar\omega^{\text{ad}}_{\mathbf{q}}$ and corrected spectrum $\hbar\omega^{\text{corr}}_{\mathbf{q}}$ plotted with fitted magnon spectrum $\hbar\omega^{\chi}_{\mathbf{q}}$}
  \label{fig:nio_chi}
\end{figure}

Nickel oxide NiO crystallizes in the cubic rocksalt structure, as shown in Fig.\ref{fig:nio_chi}(a), and is an antiferromagnetic Mott insulator with a wide band gap. The antiferromagnetic configuration consists of ferromagnetic Ni layers staked along [111] direction with alternate magnetization. The experimental cubic lattice constant $a=4.17$ \AA~\cite{wyckoff1963crystal} is used in our calculation. In order to reproduce the antiferromagnetic ground state, LDA+$U$ technique is introduced in both the DFT and DFPT calculation\cite{dudarev1998electron}, with $U$=5.4 eV for $d$-electrons of Ni\cite{dudarev1998electron,bengone2000implementation}. The resultant electronic band gap in the ground state is over 2.5 eV. 

As shown in Fig.\ref{fig:nio_chi}(d), the adiabatic spectrum $\hbar\omega^{\text{ad}}_{\mathbf{q}}$ is indeed made of two degenerate branches that display linear dispersion near $\Gamma$ as predicted by the linear spin-wave theory for antiferromagnet. Note $\hbar\omega^{\text{ad}}_{\mathbf{q}}$ already reproduces the benchmark magnon spectrum $\hbar\omega^{\chi}_{\mathbf{q}}$ accurately, and the eigenvalue refinement makes little modification (see Fig.\ref{fig:nio_chi}(c) and (d)). As the magnon modes are separated from Stoner excitations by a large gap of 2.5 eV, it is not surprising that the adiabatic theory can provide an accurate description of the spin-wave dynamics in NiO. 

\begin{figure}
  \includegraphics[width=80 mm]{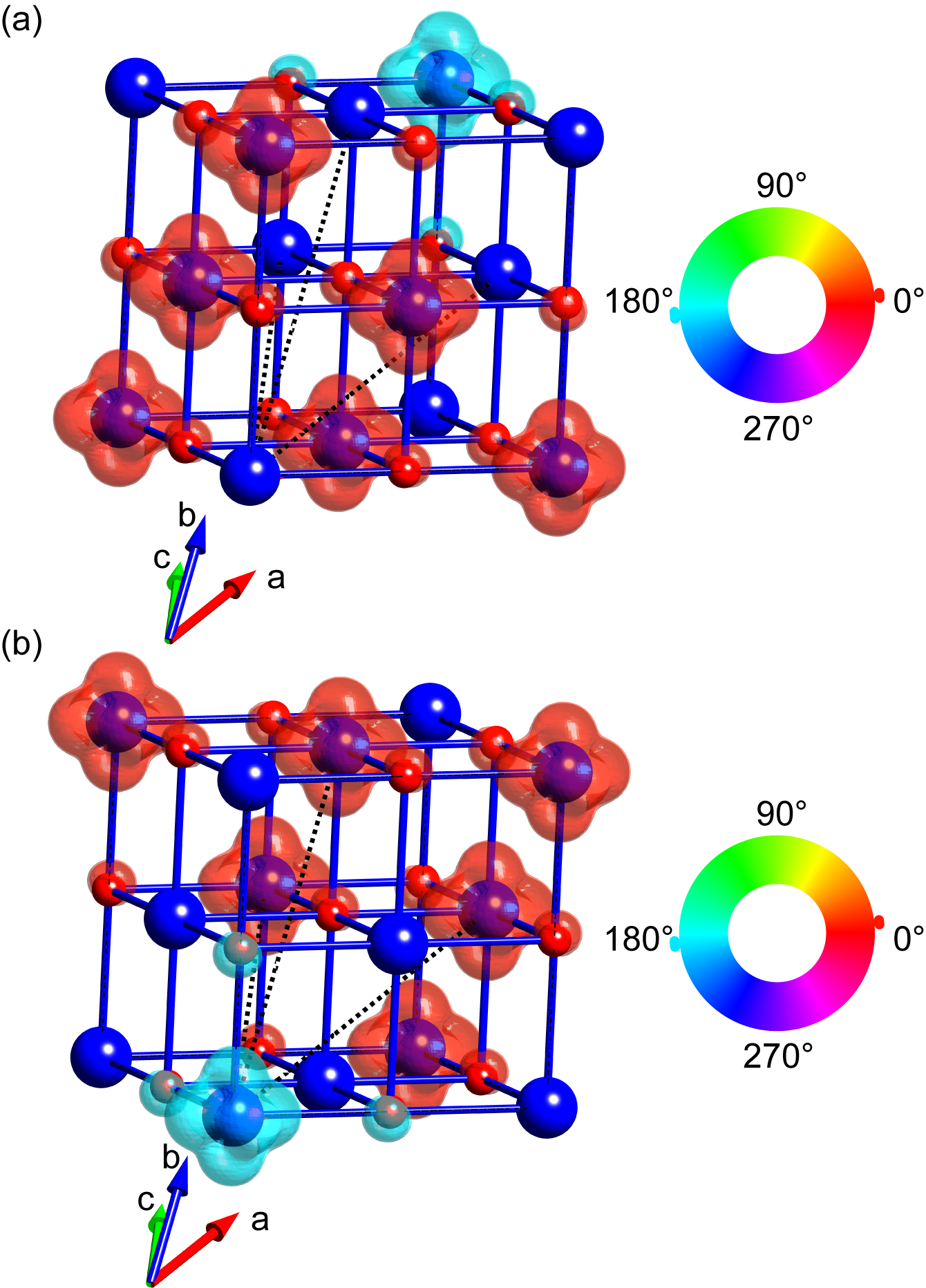}
  \caption{Magnon wavefunctions of NiO at T. (a) and (b) display isosurface of $|\delta \mathbf{m}_{\mathbf{q}}(\mathbf{r})e^{\mathrm{i}\mathbf{q}\cdot\mathbf{r}}|$ of two magnon modes, respectively. The isosurfaces are colored according to direction of Re$[\delta \mathbf{m}_{\mathbf{q}}(\mathbf{r})e^{\mathrm{i}\mathbf{q}\cdot\mathbf{r}}]$ within $x-y$ planes, whose polar angles are marked on the color wheels beside. }
  \label{fig:nio_wave}
\end{figure}

Two magnon wavefunctions of NiO at high symmetry point T ($\frac{1}{2}\mathbf{a}^*+\frac{1}{2}\mathbf{b}^*+\frac{1}{2}\mathbf{c}^*$) are plotted in Fig.\ref{fig:nio_wave} (a) and (b). Similar to CrO$_2$, spin waves in NiO also conform to the atomic moment picture, with magnetization around each Ni atom undergoes precessional motion as a rigid moment. It is also found that spin-wave dynamics in spin-up layers and spin-down layers are decoupled. This is because the precession of spins in two adjacent spin-up layers are anti-phase at crystal momentum T, and the exchange fields induced by them on the middle spin-down layer cancels. 

\subsection{bcc Fe, a non-adiabatic case\label{bccFe}}
\begin{figure}
  \includegraphics[width=80 mm]{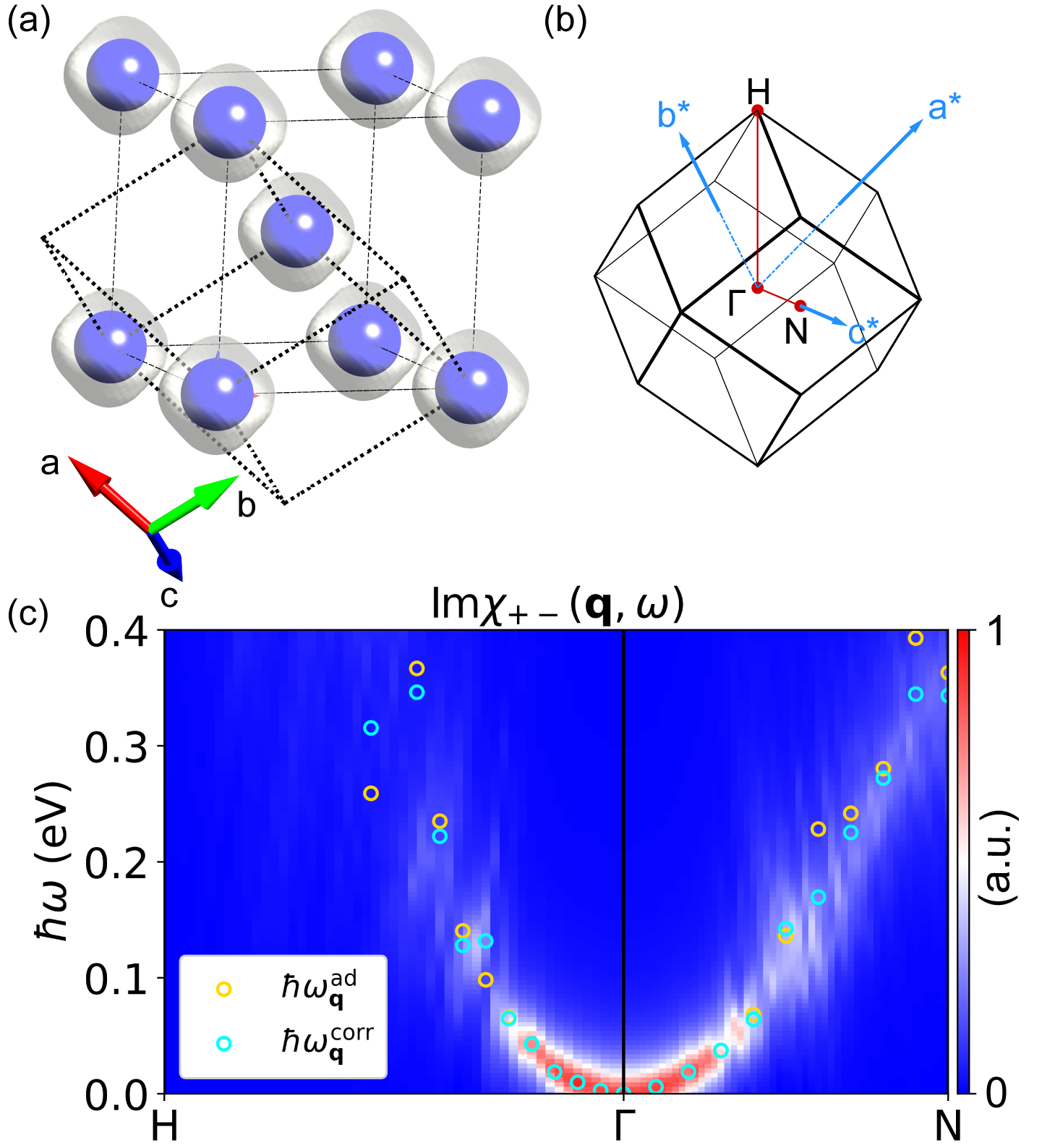}
  \caption{(a) The crystal structure of bcc Fe plotted with an isosurface of ground state $z-$magnetization. (b) The Bulk Brillouin zone with highlighted high symmetry path. (c) The spin fluctuation spectrum Im$\chi_{+-}(\mathbf{q}\omega)$. Adiabatic spectrum $\hbar\omega^{\text{ad}}_{\mathbf{q}}$ and corrected spectrum $\hbar\omega^{\text{corr}}_{\mathbf{q}}$ are also plotted.}
  \label{fig:Fe_chi}
\end{figure}

Elemental iron in body-centered cubic structure is a prototypical ferromagnet with low-lying Stoner continuum that entails severe Landau damping of magnons. We employ the experimental cubic lattice constant $a=2.86$ \AA~ \cite{wyckoff1963crystal} in our calculations, and the computed ground state magnetization $\mathbf{m}_0(\mathbf{r})$ is shown in Fig.\ref{fig:Fe_chi}(a).

The spin fluctuation spectrum Im$\chi_{+-}(\mathbf{q}\omega)$ of bcc Fe  is shown in Fig.\ref{fig:Fe_chi}(c). Magnon resonance peaks are evident only within $|\mathbf{q}|\sim 0.2$ $\Gamma\text{N}$, beyond which they get significantly attenuated due to a overlapping Stoner continuum. The Landau damping of magnon results in severe non-adiabaticity, and the adiabatic EOM is not expected to provide a good description of spin waves. But we still apply our algorithm to bcc Fe to inspect its behavior in such
non-adiabatic cases. It is found, if the basis expansion is conducted, the energy of lowest eigenmode beyond $|\mathbf{q}|\sim 0.2$ $\Gamma\text{N}$ can be quite close to zero. Such an eigenmode indeed corresponds to an low-lying Stoner excitation, which also satisfies the exact EOM Eq.\eqref{exact}, with the electron-hole pair from majority-spin and minority-spin Fermi surfaces. These low-lying Stoner eigenmodes will not appear if we stick to the initial set of Gaussian fields and the resultant adiabatic spectrum $\hbar\omega^{\text{ad}}_{\mathbf{q}}$ is plotted in Fig.\ref{fig:Fe_chi}(c), along with the corrected spectrum $\hbar\omega^{\text{corr}}_{\mathbf{q}}$.

The adiabatic EOM predicts an eigenmode with eigenvalue higher than 1 eV near H ($\frac{1}{2}a^*+\frac{1}{2}b^*-\frac{1}{2}c^*$), 
which is deep inside the Stoner continuum and unlikely a magnon mode. $\hbar\omega^{\text{ad}}_{\mathbf{q}}$ already accurately predicts the magnon energy position within $|\mathbf{q}|\sim 0.2$ $\Gamma\text{N}$,
 and the eigenvalue refinement makes little modification to the spectrum. But beyond $|\mathbf{q}|\sim 0.2$ $\Gamma\text{N}$, $\hbar\omega^{\text{ad}}_{\mathbf{q}}$ and $\hbar\omega^{\text{corr}}_{\mathbf{q}}$ 
depart from each other, and the latter generally lie closer to the spectral maximum in Im$\chi_{+-}(\mathbf{q}\omega)$. This is a property of the eigenvalue refinement that the predicted eigenvalue will approach the energy of the state making dominant contribution to magnetization response, and in each iteration the contribution of this state is further enhanced by near-resonance driving fields.

\subsection{Na$_2$IrO$_3$ and the breakdown of atomic moment picture.\label{nairo}}

\begin{figure}
  \includegraphics[width=80 mm]{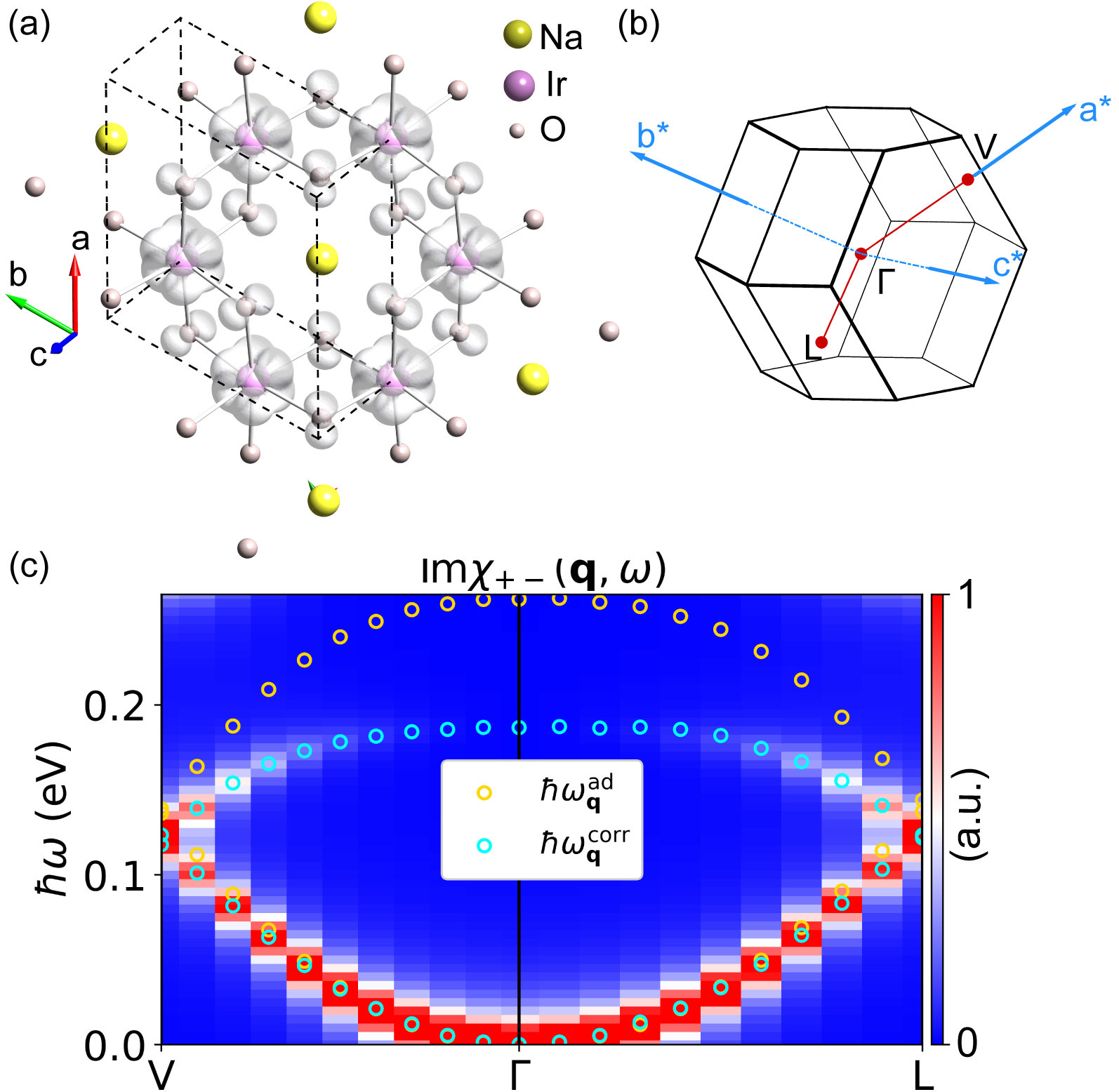}
  \caption{(a) The crystal structure of Na$_2$IrO$_3$ plotted with an isosurface of ground state $z-$magnetization. (b) The Bulk Brillouin zone with highlighted high symmetry path. (c) The spin fluctuation spectrum Im$\chi_{+-}(\mathbf{q}\omega)$. Adiabatic spectrum $\hbar\omega^{\text{ad}}_{\mathbf{q}}$ and corrected spectrum $\hbar\omega^{\text{corr}}_{\mathbf{q}}$ are also plotted.}
  \label{fig:nairo_chi}
\end{figure}

Sodium iridate Na$_2$IrO$_3$ contains layered honeycomb lattices of Ir atoms, as shown in Fig.\ref{fig:nairo_chi}(a). Each Ir atom is surrounded by an oxygen octahedron with trigonal contraction. An Ir$^{4+}$ ion has a $d^5$ electronic configuration occupying the $t_{2g}$ manifold in the cubic crystal field, which effectively exhibits spin$-\frac{1}{2}$ and orbital angular momentum $L=1$\cite{Khonskii2014transition}. It was argued that an Ir$^{4+}$ with SOC can produce an effective $j_{\text{eff}}=1/2$ pseudospin and magnetic phenomena in Na$_2$IrO$_3$ is typically explained by a $j_{\text{eff}}=1/2$ Kitaev-Heisenberg spin Hamiltonian\cite{jackeli2009mott,chaloupka2010kitaev}. However, sophisticated terms like off-diagonal spin iteration and longer-range spin interactions often have to be included for quantitative fitting of experimental data\cite{rau2014generic,kim2020dynamic,liu2022exchange}.

It was suggested Na$_2$IrO$_3$ is nearly itinerant, as its $5d$ bandwidth (1.5-2 eV) is comparable to the Hubbard on-site repulsion $U$(1-2 eV)\cite{mazin20122,foyevtsova2013ab,mazin2013}.  Moreover, low-energy electronic structure of Na2IrO3 is characterized by the formation of quasi-molecular orbitals where 
a $t_{2g}$ electron is localized on an Ir$_6$ hexagon by directional oxygen-mediated hoppings but fully delocalized over six Ir sites\cite{mazin20122,foyevtsova2013ab,mazin2013}. As the three $t_{2g}$ orbitals of a single Ir atom separately belongs to tightly-locked molecular orbitals on three different Ir$_6$ hexagons that share this Ir atom, they may not behave coherently as a rigid atomic moment in spin-wave dynamics. To see whether this happens requires {\it ab initio} calculation of magnons in Na$_2$IrO$_3$. 

We study the Na$_2$IrO$_3$ without SOC in its ground state ferromagnetic phase. As an inconsequential but convenient simplification, we use an idealized structure of Na$_2$IrO$_3$, dubbed S1 in Ref.\cite{foyevtsova2013ab}, removing orthorhombic distortions to restore the $D_{6h}$ symmetry of Ir hexagons with equal nearest-neighbor Ir-O bonds. The electronic structure in this idealized structure is sufficiently close to that from the experimental structure, and is characterized by formation of quasi-molecular orbitals\cite{foyevtsova2013ab}. LDA+$U$ technique is introduced in both the DFT and DFPT calculations using $U=3.8$ eV for $d$-electrons of Ir, yielding a Stoner gap $\sim300$ meV. (see Appendix.\ref{AppB})

The computed spin fluctuation spectrum Im$\chi_{+-}(\mathbf{q}\omega)$ is shown in Fig.\ref{fig:nairo_chi}(c). While there are two branches of magnon as the primitive cell has two magnetic Ir atoms, the resonance peak of the optical magnon gradually fades away near $\Gamma$. This indicates the optical magnon can barely be excited by an atom-centered Gaussian transverse field, which is already hinting at the failure of the atomic moment picture.

Indeed, Na$_2$IrO$_3$ is an extraordinary example that necessitates our iterative procedure to capture its spin-wave configurations. In the cases of CrO$_2$ and NiO, the adiabatic EOM Eq.\eqref{generalized} in the initial subspace made of atomic Gaussian fields already produces $\hbar\omega^{\text{ad}}_{\mathbf{q}}$ converged within 1 meV. However, for Na$_2$IrO$_3$, the iterative eigensolver fails to optimize the upper eigenvalue near $\Gamma$ with such an initial subspace, as atomic Gaussian fields can barely excite the optical magnon there. Remarkably, the two magnon modes at BZ boundary can still be solved by our procedure after a few iterations, and the eigenfields $\{\mathbf{B}^{\alpha}_{\mathbf{q}}\}$ solved there turn out to form a good initial subspace at $\Gamma$. 
The eigenfields solved at $\Gamma$ are in turn used as the initial subspace at other momentum. The resultant adiabatic spectrum $\hbar\omega^{\text{ad}}_{\mathbf{q}}$ and the corrected spectrum $\hbar\omega^{\text{corr}}_{\mathbf{q}}$ are also plotted in Fig.\ref{fig:nairo_chi}(c). It is found that $\hbar\omega^{\text{corr}}_{\mathbf{q}}$ accurately reproduces the energies of magnon resonance peaks for both the acoustic and optical branch.

\begin{figure}
  \includegraphics[width=80 mm]{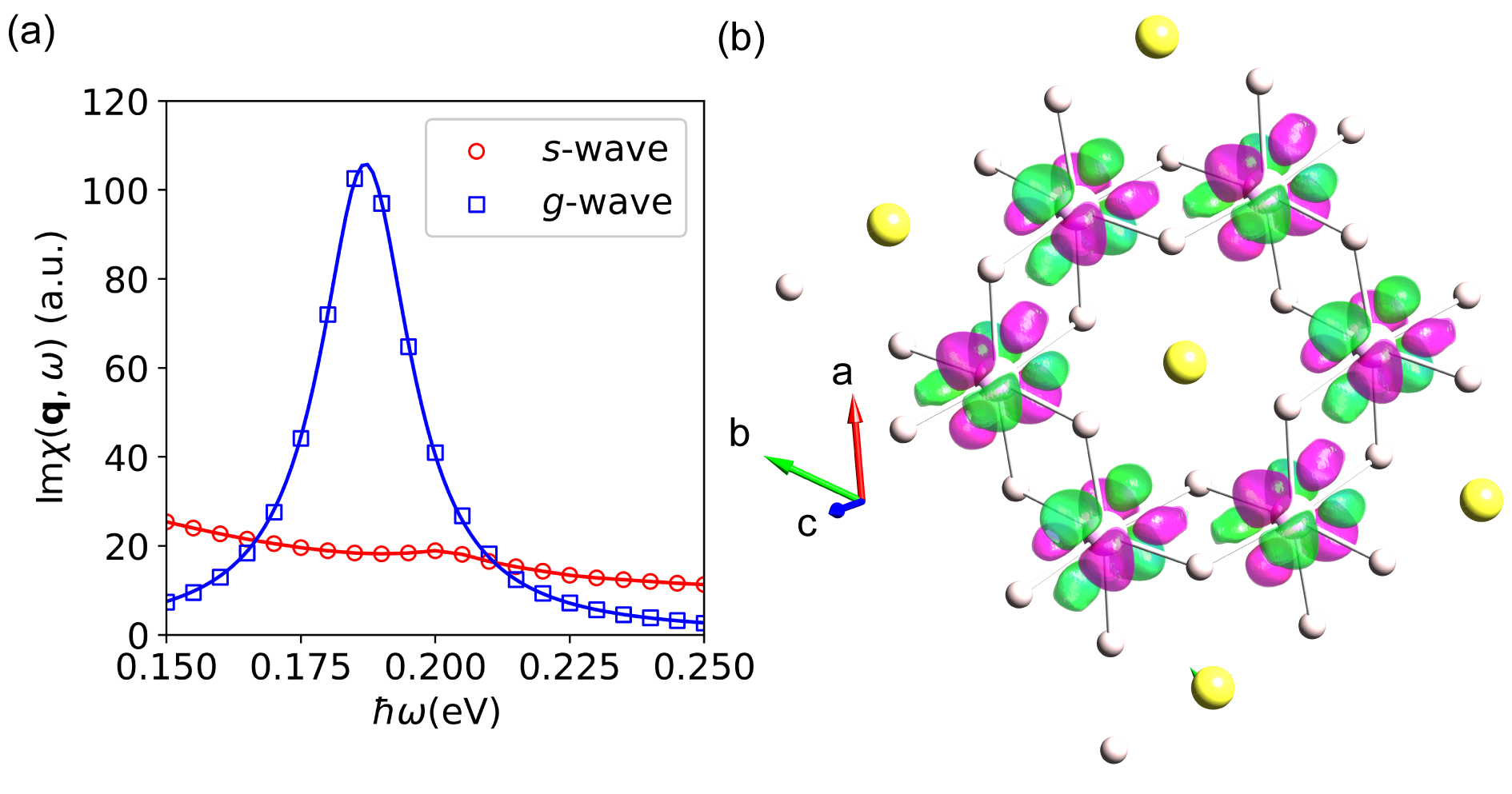}
  \caption{(a) The spin fluctuation spectrum at $\Gamma$ near the optical branch of magnon. We have used two kinds of excitation field $\mathbf{B}_{\mathbf{q}}(\mathbf{r})$, the $s$-wave Gaussian field as before and the eigenfield for the optical branch that manifests $g$-wave character. The plotted functions are Im$\chi(\mathbf{q}\omega)$=$\int d\mathbf{r} \mathbf{B}^*_{\mathbf{q}}(\mathbf{r})\cdot\mathbf{m}_{\mathbf{q}}(\mathbf{r}\omega)$ in each case. The $g$-wave signal has been fitted using Lorentz lineshape.(b) Isosurface of the $g$-wave field $|\delta \mathbf{B}_{\mathbf{q}}(\mathbf{r})e^{\mathrm{i}\mathbf{q}\cdot\mathbf{r}}|$,  colored according to direction of Re$[\delta \mathbf{B}_{\mathbf{q}}(\mathbf{r})e^{\mathrm{i}\mathbf{q}\cdot\mathbf{r}}]$ within $x-y$ planes.}
  \label{fig:nairo_gamma}
\end{figure}

\begin{figure*}
  \includegraphics[width=140 mm]{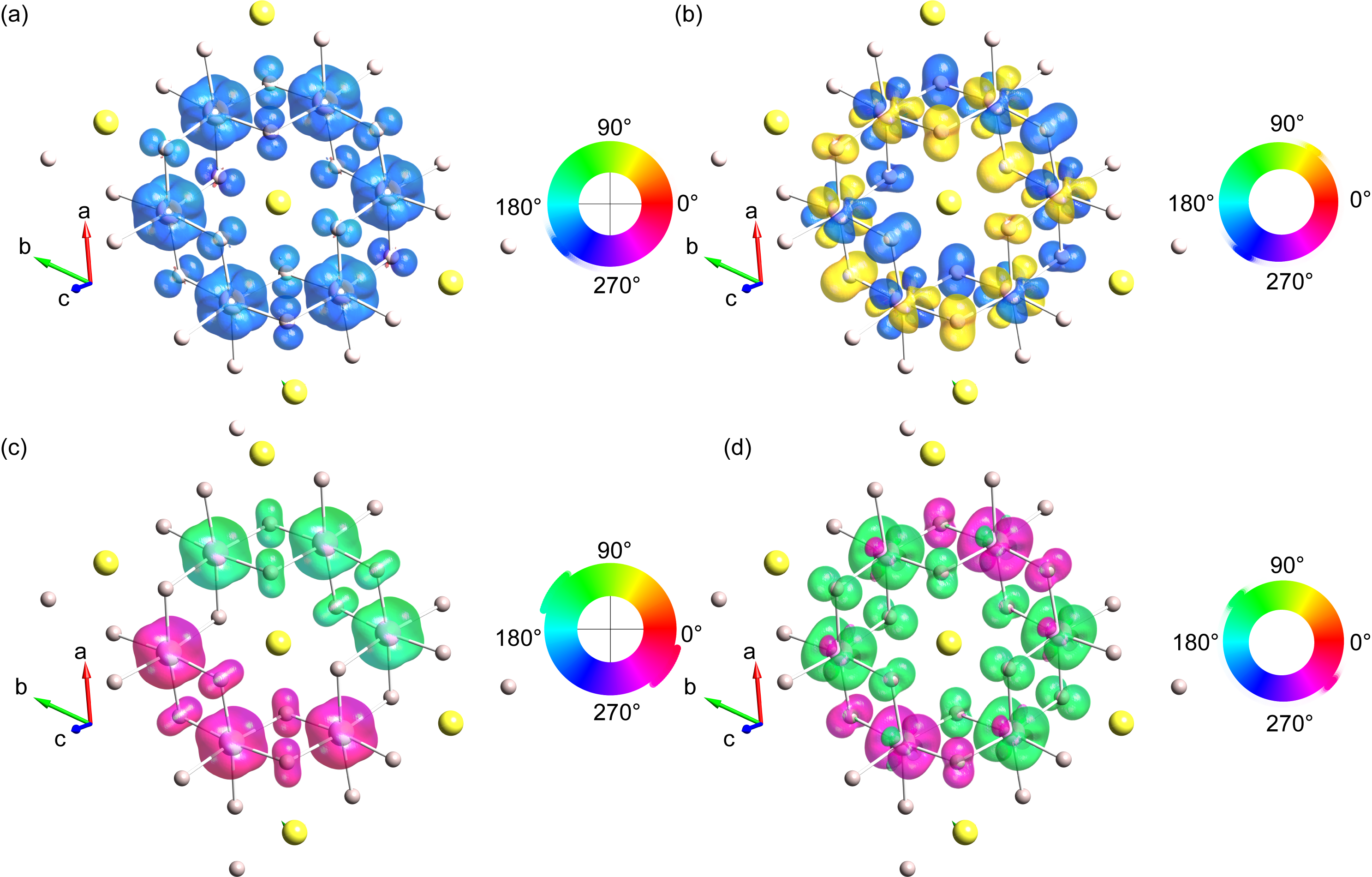}
  \caption{(a) and (b) Magnon wavefunctions of Na$_2$IrO$_3$ at $\Gamma$. (c) and (d) Magnon wavefunctions of Na$_2$IrO$_3$ at V. Isosurface of $|\delta \mathbf{m}_{\mathbf{q}}(\mathbf{r})e^{\mathrm{i}\mathbf{q}\cdot\mathbf{r}}|$ of two magnon modes are displayed, respectively. The isosurfaces are colored according to direction of Re$[\delta \mathbf{m}_{\mathbf{q}}(\mathbf{r})e^{\mathrm{i}\mathbf{q}\cdot\mathbf{r}}]$ within $x-y$ planes, whose polar angles are marked on the color wheels beside. }
  \label{fig:nairo_wave}
\end{figure*}

We then investigate the magnon wavefunctions of Na$_2$IrO$_3$, as shown in Fig.\ref{fig:nairo_wave}(a) and (b) at $\Gamma$ and Fig.\ref{fig:nairo_wave}(c) and (d) at V. The transverse spin fluctuation due to acoustic magnon are unidirectional around each Ir atom and have spatial profile similar to that of the ground state magnetization in the Fig. \ref{fig:nairo_chi}(a), more or less conforming to the atomic moment picture. On the other hand, the transverse spin fluctuation of optical magnons are characterized by anti-parallel lobes, which does not align with atomic moment picture. These magnon wavefunctions intuitively explain the inactivity of the optical magnon at $\Gamma$ under atomic Gaussian transverse field, since an $s$-wave field on a magnetic atom could not generate response with higher angular momenta. Indeed, we find an external field with $g$-wave character (orbital angular momentum $l=4$) on Ir sites is needed to excite the optical magnon at $\Gamma$, as shown in Fig.\ref{fig:nairo_gamma}.

Further quantitative insights in the breakdown of the atomic moment picture may be gained by inspecting the density matrices. We compute the ground state single-particle density matrix $\varrho^0$ on an Ir site,
\begin{equation}
  \varrho^0_{m\sigma,n\sigma'}=\langle \Phi_0|d^{\dagger}_{m\sigma}d_{n\sigma'}|\Phi_0\rangle,
\end{equation}
and its fluctuation $\delta\varrho^{\alpha}$ after excitation of magnon $\hbar\omega_{\mathbf{q}}^{\alpha}$ ($\alpha=1,2$ for acoustic and optical magnon, respectively),
\begin{equation}
  \delta\varrho^{\alpha}_{m\sigma,n\sigma'}=\langle \Phi_0|d^{\dagger}_{m\sigma}d_{n\sigma'}|\delta\Phi^{\alpha}_{\mathbf{q}}\rangle+\langle\delta \Phi^{\alpha}_{-\mathbf{q}}|d^{\dagger}_{m\sigma}d_{n\sigma'}|\Phi_0\rangle,
\end{equation}
where $d^{\dagger}_{m\sigma}$ ($d_{m\sigma}$) is the creation (annihilation) operator of a $d$-orbital with spin $\sigma=\uparrow,\downarrow$ indexed by $m,n=xy,yz,z^2,xz,x^2-y^2$. Define decomposition of a spinor density matrix,
$\varrho_{m\sigma,n\sigma'}=\sum_i \varrho^i_{mn}\otimes \sigma^i_{\sigma\sigma'}$,
where $\sigma^i$ are Pauli matrices with $i=x,y,z$ for magnetization channels and the identity matrix with $i=0$ for the charge channel. A principal component analysis reveals that each density matrices in the magnetization channels shares three major components, $\varrho^{\text{I}}$,$\varrho^{\text{II}}$ and $\varrho^{\text{III}}$. Their proportions in each density matrices are displayed in Fig.\ref{fig:PCA}(a), and their matrix representations in the bases of effective orbital angular momentum $L_Z$ (see Appendix \ref{AppC}) are shown in Fig.\ref{fig:PCA}(b),\ref{fig:PCA}(c) and \ref{fig:PCA}(d), respectively. 

Apparently, $\varrho^{0z}$ and $\delta\varrho^{1x,y}$ shares a single dominant component $\varrho^{\text{I}}$ that is diagonal in the $L_Z$ basis and occupies mostly $L_Z=\pm 1$ states. Thus, the on-site density matrix $\varrho$ during acoustic spin-wave dynamics is of the form (with real coefficient $\mu$ and $\nu$.)
\[\varrho(t)=\varrho^{\text{I}}\otimes[\mu\sigma_z+\nu(\sigma_x\cos\omega t+\sigma_y\sin\omega t),\]
which describes a rigid precessing moment whose spatial profile is specified by the same $\varrho^{\text{I}}$ as the ground state magnetization. This is exactly what the atomic moment picture implies. At the same time, the first-order wavefunction at $\Gamma$ satisfies 
\[|\delta\Phi^{\alpha}_{\pm\mathbf{q}}\rangle\propto \sum_{\mathbf{R}} S^{\text{at}}_{-}(\mathbf{R})|\Phi_0\rangle,\]
where $S^{\text{at}}_{-}(\mathbf{R})=\sum_m d^{\dagger}_{m\downarrow}(\mathbf{R})d_{m\uparrow}(\mathbf{R})$ stand for the ladder operator of an atomic spin and $\mathbf{R}$ runs over all Ir sites.
However, for the optical magnon, $\delta\varrho^{2x,y}$ are made of two other components $\varrho^{\text{II}}$ and $\varrho^{\text{III}}$ that are mainly off-diagonal in the $L_Z$ basis, indicating on-site interorbital transition. The on-site density matrix $\varrho$ during optical spin-wave dynamics instead will be of the form (with real coefficient $\mu$, $\varepsilon$ and $\kappa$.)
\[
\begin{aligned}
\varrho(t)=\mu\varrho^{\text{I}}\otimes\sigma_z+&
\epsilon \varrho^{\text{II}}\otimes (-\sigma_x\sin\omega t+\sigma_y\cos\omega t)\\
+&
\kappa \varrho^{\text{III}}\otimes (\sigma_x\cos\omega t+\sigma_y\sin\omega t),
\end{aligned}
\]
which indicates that the transverse magnetization not only has a different spatial profile from the longitudinal magnetization, but is generally non-collinear with two precessing components with a phase lag. These 
clearly signals the failure of the atomic moment picture to describe the optical magnon in Na$_2$IrO$_3$. Given the formation of molecular orbitals in the ground state, the elemental spin DOF may be better described in terms of these molecular orbitals, $S^{\text{mol}}_{-}(\mathbf{R})=\sum_{\mathcal{M},\mathcal{M}'} c_{\mathcal{M}\mathcal{M}'} d^{\dagger}_{\mathcal{M}\downarrow}(\mathbf{R})d_{\mathcal{M}'\uparrow}(\mathbf{R})$, with $\mathbf{R}$ runs over all Ir$_6$ hexagons and $\mathcal{M}$,$\mathcal{M}'$ over six kinds of molecular orbitals\cite{mazin20122}. $c_{\mathcal{M}\mathcal{M}'}=\delta_{\mathcal{M}\mathcal{M}'}$ will corresponds to a rigid molecular moment picture. However, this molecular moment picture alone cannot provide a coherent description of both acoustic and optical magnons in Na$_2$IrO$_3$, since one single molecular moment per unit cell cannot account for two branches of magnons. To rigorously identify the elemental spin DOF in Na$_2$IrO$_3$ as magnons are concerned, may require wannierization the computed magnon wavefunctions.


\begin{figure}
  \includegraphics[width=80 mm]{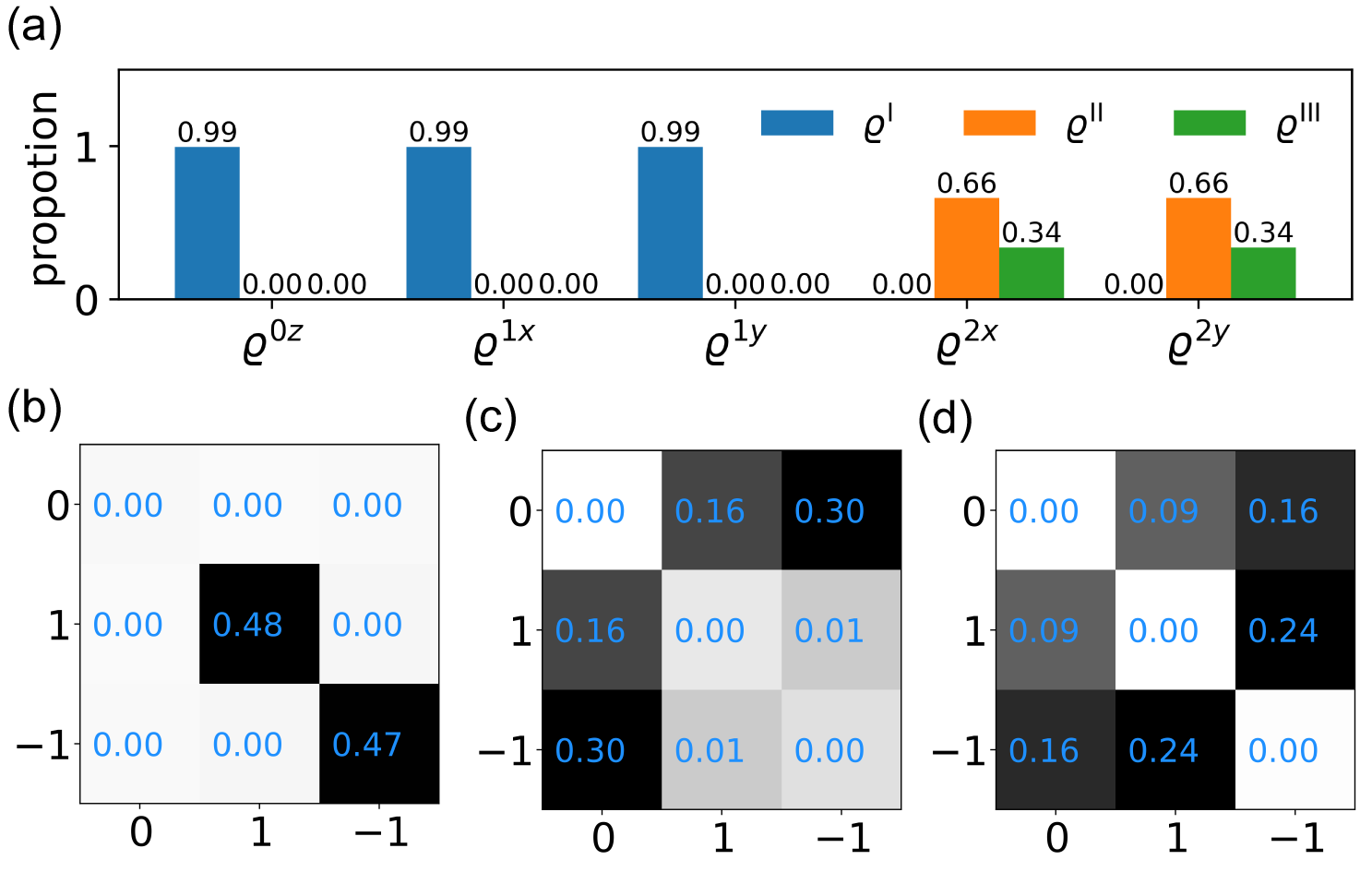}
  \caption{(a) Proportion of major principal components in each density matrices. (b),(c) and (d) Matrix $|\langle L_Z|\varrho|L'_Z\rangle|$ with $\varrho=\varrho^{\text{I}},\varrho^{\text{II}}$ and $\varrho^{\text{III}}$, respectively. Bases are given in the order $L_Z$=0,1 and -1.}
  \label{fig:PCA}
\end{figure}
 
\section{Conclusions\label{sec:conclusion}}

To summarize, we have developed a new model-free method for {\it ab initio} calculation of magnon wavefunction in real materials. This method is based on Niu-Kleinman adiabatic EOM for spin-wave dynamics but supplemented by several reformulations and extensions to overcome the need for atomic moment ansatz in numerical solutions and the imperfect adiabaticity at higher energy, and enabled by our previous implementation of DFPT. It also achieves more efficient and systematic evaluation of magnon spectrum than calculation of spin fluctuation spectrum in systems without significant Landau damping. Our method provides a first principles description of collective spin fluctuations in real materials, beyond the usual spin Hamiltonian approaches that presuppose atomic moment picture.

We have implemented this method using the DFPT within the PAW framework\cite{liu2023implementation} and  computed the {\it ab initio} magnon wavefunctions for a few typical and not-so-typical magnetic materials, in order to assess the validity of the atomic moment picture in describing collective spin fluctuations. The results for ferromagnetic CrO$_2$ and antiferromagnet NiO are in perfect agreement with the atomic moment picture. In their magnon dynamics, the magnetization densities around a magnetic atom remain rigid and unidirectional. Our results on magnon wavefunctions of Na$_2$IrO$_3$ in the optical branch reveals the breakdown of the atomic moment picture on the level of td mean field theory. This is likely to be a consequence of formation of quasi-molecular orbitals in the electronic structure and orbital locking due to inter-atomic hopping outplays the Hund's exchange coupling, which is also relevant when SOC is included\cite{foyevtsova2013ab}. 

The breakdown of the atomic moment picture also challenges the use of effective spin models Eq.(1) to investigate magnetic phenomena. Insisting on the spin model approach for a coherent interpretation of various magnetic phenomena in a material can leads to overfitting a very sophisticated Hamiltonian. Instead, we may need a revised notion of low energy DOF of spin fluctuations to incorporate the breakdown of the atomic moment picture in spin-wave dynamics. Such a revision will be very important to the field of spin waves, as most theoretical consideration including recent topics like symmetry classification of magnon band\cite{corticelli2022spin,chen2023spin} and magnon-phonon coupling\cite{wang2023}, are mostly based on spin Hamiltonian approaches. Our method provides a concrete computational assay for physically sensible models that capture the true low energy DOF in magnetic dynamics. Our results also suggest experimental signatures for the breakdown of the atomic moment picture. As illustrated in the exemplifying Na$_2$IrO$_3$ , the spin fluctuations of the optical magnon are of higher angular momentum ($l$=4) with no net dipole moment on each magnetic Ir site. These modes then will have diminishing cross section in inelastic neutron scattering that relies on magnetic dipole-dipole interaction, leading spectroscopic features similar to Fig.\ref{fig:nairo_chi}(c). On the other hand, the resonant inelastic X-ray scattering, taking advantage of orbital-selective transition, can exert an effective anisotropic field on Ir sites and may be able to detect such highly anisotropic spin fluctuations. A comparison of this sort will then an experimental assay for the breakdown of the atomic picture in the real materials.

\begin{acknowledgments}
  We acknowledge the financial support from the National Natural Science Foundation of China (Grant Nos. 12274003, 11725415 and 11934001), the National Key R\&D Program of China (2021YFA1400100), and the Innovation Program for Quantum Science and Technology (Grant No. 2021ZD0302600). 
  We are grateful for informative discussions with Yuan Li as we were finalizing the manuscript. 
\end{acknowledgments}

\appendix
\section{The adiabatic EOM with finite $\eta$\label{AppA}}
DFPT calculations are usually performed with a finite switch-on parameter $\eta$. It is needed to remove the singularity in Eq.\eqref{Psi} due to zero denominators, which will otherwise leads to numerically non-convergent principal value integrals in the calculation of induced densities and divergent response at magnon resonance. 
The perturbated wavefunction $|\delta\Psi _{\pm\mathbf{q}}(\pm\omega)\rangle$ formally expands as 
\begin{equation}
  |\delta\Psi _{\pm\mathbf{q}}(\pm\omega)\rangle = \sum_n \frac{|\Psi_{n\pm\mathbf{q}}\rangle\langle\Psi_{n\pm\mathbf{q}}|O_{\pm\mathbf{q}}|\Psi_{0}\rangle}{\pm\omega+E_0-E_{n\pm\mathbf{q}}+\mathrm{i}\eta},\label{Psi}
\end{equation}
with $O_{\mathbf{q}}=\int d\mathbf{r} \hat{\mathbf{m}}(\mathbf{r})\cdot \mathbf{B}_{q}(\mathbf{r})e^{\mathrm{i}\mathbf{q}\cdot\mathbf{r}}$ and $O_{-\mathbf{q}}\equiv O^{\dagger}_{\mathbf{q}}$. 
Spin susceptibility in Eq.\eqref{chi} formally expands as 
\begin{equation}
  \begin{aligned}
    \chi^{\eta}_{ij}(\mathbf{r},\mathbf{r}',\omega)=\sum_n -&\frac{\langle \Psi_0|\hat{m}_i(\mathbf{r})|\Psi_n\rangle \langle \Psi_n|\hat{m}_j(\mathbf{r}')|\Psi_0\rangle}{\omega +E_0 -E_n +\mathrm{i}\eta }\\
    &-\frac{\langle \Psi_0|\hat{m}_j(\mathbf{r}')|\Psi_n\rangle \langle \Psi_n|\hat{m}_i(\mathbf{r})|\Psi_0\rangle}{-\omega +E_0 -E_n -\mathrm{i}\eta }, 
  \end{aligned}
\end{equation}
where we explicitly distinguish $\chi(\omega)$'s calculated with different $\eta$ by their labels. It is apparent that $\chi^{\eta}(\omega)$ is related to $\chi(\omega)\equiv\chi^{0^+}(\omega)$ by an analytical continuation, $\chi^{\eta}(\omega)=\chi(\omega+i\eta)$. Thus, a pole of $\chi(\omega)$ along the real $\omega$ axis will appear as an resonance peak in $\chi^{\eta}(\omega)$ at the same (real) frequency but with broadening $\eta$. The exact EOM Eq.\eqref{exact} in term of $\chi^{\eta}$ writes
\begin{equation}
  \sum_j[\chi^{\eta}(\omega-i\eta)^{-1}]_{ij}\delta m_j=0.
\end{equation} 
By substitution of $\chi^{\eta}(\omega-i\eta)^{-1}$ by its truncated low frequency expansion,
\begin{equation}
  [\chi^{\eta}(\omega-i\eta)]^{-1}\approx[\chi^{\eta}(0)]^{-1}+(\omega-i\eta)\frac{\partial}{\partial \omega}[\chi^{\eta}(0)]^{-1}, 
\end{equation}
we find the following EOM of $ \mathbf{B}=\chi^{\eta}(0)^{-1}\delta \mathbf{m}$ at finite $\eta$,
\begin{equation}
  \sum_{j}\chi_{ij}^{\eta}(0)B_j=(\omega-i\eta)\sum_{j}\Big[\frac{\partial}{\partial \omega}\chi^{\eta}(0)\Big]_{ij}B_j. \label{A4}
\end{equation} 
The matrix representation $S(\omega)$ of $\chi^{\eta}(\omega)$ can be calculated using Eq.\eqref{Sab} as before, while the matrix representation $H(\omega)$ of $\frac{1}{\hbar}\frac{\partial}{\partial \omega}\chi^{\eta}(\omega)$ rigorously satisfies the following identity 
\begin{equation}
\begin{aligned}
  H_{ab}&(\omega)=\\
&\langle\delta\Psi_{\mathbf{q}}^{-\eta,a}(\omega)|\delta\Psi_{\mathbf{q}}^{\eta,b}(\omega)\rangle-\langle\delta\Psi_{-\mathbf{q}}^{\eta,b}(-\omega)|\delta\Psi_{-\mathbf{q}}^{-\eta,a}(-\omega)\rangle,  \label{A5}
\end{aligned}
\end{equation}
where we have further distinguished distinguish $|\delta\Psi_{\pm\mathbf{q}}(\pm \omega)\rangle$'s calculated with different $\eta$ by their labels.

The spin susceptibility $\chi^{\eta}$ calculated from DFPT under the adiabatic DFT approximation satisfies
\begin{equation}
  \chi^{\eta}(\omega)^{-1}=\chi^{\eta}_0(\omega)^{-1}-f,
\end{equation}
thus $\frac{\partial}{\partial \omega}\chi^{\eta}$ satisfies 
\begin{equation}
\begin{aligned}
  \frac{\partial}{\partial \omega}&\chi^{\eta}(\omega)=\\
  &\left[ \chi_{0}^{-\eta }( \omega )^{-1} \chi ^{-\eta }( \omega )\right]^{\dagger }\left[\frac{\partial }{\partial \omega } \chi _{0}^{\eta }( \omega )\right] \chi _{0}^{\eta }( \omega )^{-1} \chi ^{\eta }(\omega), 
\end{aligned}\label{A7}
\end{equation}
where we have used $\chi^{\eta}(\omega)=\chi^{-\eta}(\omega)^{\dagger}$. Note $\chi _{0}^{\pm\eta }( \omega )^{-1} \chi ^{\pm\eta }(\omega)$ map the external field $B$ to self-consistent variation of Kohn-Sham field $B_{\mathrm{KS}}$, and Eq.\eqref{A7} proves that matrix representation of $\frac{1}{\hbar}\frac{\partial}{\partial \omega}\chi^{\eta}$ can be calculated using Eq.\eqref{A5} with perturbated wavefunction $|\delta\Psi\rangle$ replaced by their DFPT counterpart $|\delta \Phi\rangle$. Therefore, solution of the adiabatic EOM at finite $\eta$ in principle can proceed as that at $\eta=0$ discussed in the main text in the DFPT framework.  

In practice, calculation based on Eq.\eqref{A5} has the disadvantage that self-consistent DFPT calculation should be performed at both $+\eta$ and $-\eta$, doubling the computational burden. It is possible to find a way to avoid this if we inspect the adiabatic EOM Eq.\eqref{A4} more closely. Again, we expand it using state-resolved spin fluctuation
\begin{equation}
  \delta \mathbf{m}_{n\pm\mathbf{q}}(\mathbf{r})=
-\frac{\langle \Psi_0|\hat{\mathbf{m}}(\mathbf{r})|\Psi_{n\pm\mathbf{q}}\rangle\langle\Psi_{n\pm\mathbf{q}}|O_{\pm\mathbf{q}}|\Psi_0\rangle}{\pm\omega+E_0-E_n+i\eta},
\end{equation}
at $\omega=0$,
which leads to the following equation
\begin{widetext}
\begin{equation}
    \frac{1}{\omega-i\eta}\Big[\sum_n \delta\mathbf{m}_{n\mathbf{q}}+\sum_m\delta\mathbf{m}^*_{m\mathbf{-q}})\Big]=\sum_n \frac{1}{E_{n\mathbf{q}}-E_0-i\eta}\delta\mathbf{m}_{n\mathbf{q}}+\sum_m\frac{1}{E_0-E_{m-\mathbf{q}}-i\eta}\delta\mathbf{m}^*_{m\mathbf{-q}}.    
\end{equation}  
\end{widetext}
An low-energy eigenfield $\mathbf{B}^{\alpha}_{\mathbf{q}}$ of the adiabatic EOM should selectively excite an magnon state, so that one $\delta\mathbf{m}_{n\mathbf{q}}$ or $\delta\mathbf{m}^*_{m\mathbf{-q}}$ dominates over other terms. In this case,  
\begin{widetext}
\begin{equation}
\langle\delta\Psi_{\mathbf{q}}^{\eta,\alpha}|\delta\Psi_{\mathbf{q}}^{\eta,\beta}\rangle-\langle\delta\Psi_{-\mathbf{q}}^{\eta,\beta}|\delta\Psi_{-\mathbf{q}}^{\eta,\alpha}\rangle\\
  \approx \delta_{\alpha\beta}\frac{\omega_{\alpha}-i\eta}{\omega_{\alpha}+i\eta}\Big[\langle\delta\Psi_{\mathbf{q}}^{-\eta,\alpha}|\delta\Psi_{\mathbf{q}}^{\eta,\beta}\rangle-\langle\delta\Psi_{-\mathbf{q}}^{\eta,\beta}|\delta\Psi_{-\mathbf{q}}^{-\eta,\alpha}\rangle\Big]
\end{equation}
\end{widetext}
with $\omega_{\alpha}=E_{n\mathbf{q}}-E_0$ or $E_0-E_{m-\mathbf{q}}$. Thus, we can instead solve the adiabatic EOM 
\begin{equation}
  \hbar(\omega+i\eta)\sum_{b} H_{ab}\lambda_{b}=\sum_b S_{ab}\lambda_{b},
\end{equation}
with $H$ calculated according to Eq.\eqref{Hab} but with $|\delta\Psi_{\pm\mathbf{q}}\rangle$ replaced by $|\delta \Psi^{\eta}_{\pm\mathbf{q}}\rangle$. This modification will be insignificant when $\eta$ is small compared with magnon energy $\omega_{\alpha}$, and accurate when $\omega_{\alpha}$ is small and magnon contribution dominates the magnetization response, leading to satisfactory accuracy over the whole magnon spectrum.

\section{Formation of quasi-molecular orbitals in ferromagnetic Na$_2$IrO$_3$ \label{AppB}}
\begin{figure}
  \includegraphics[width=80 mm]{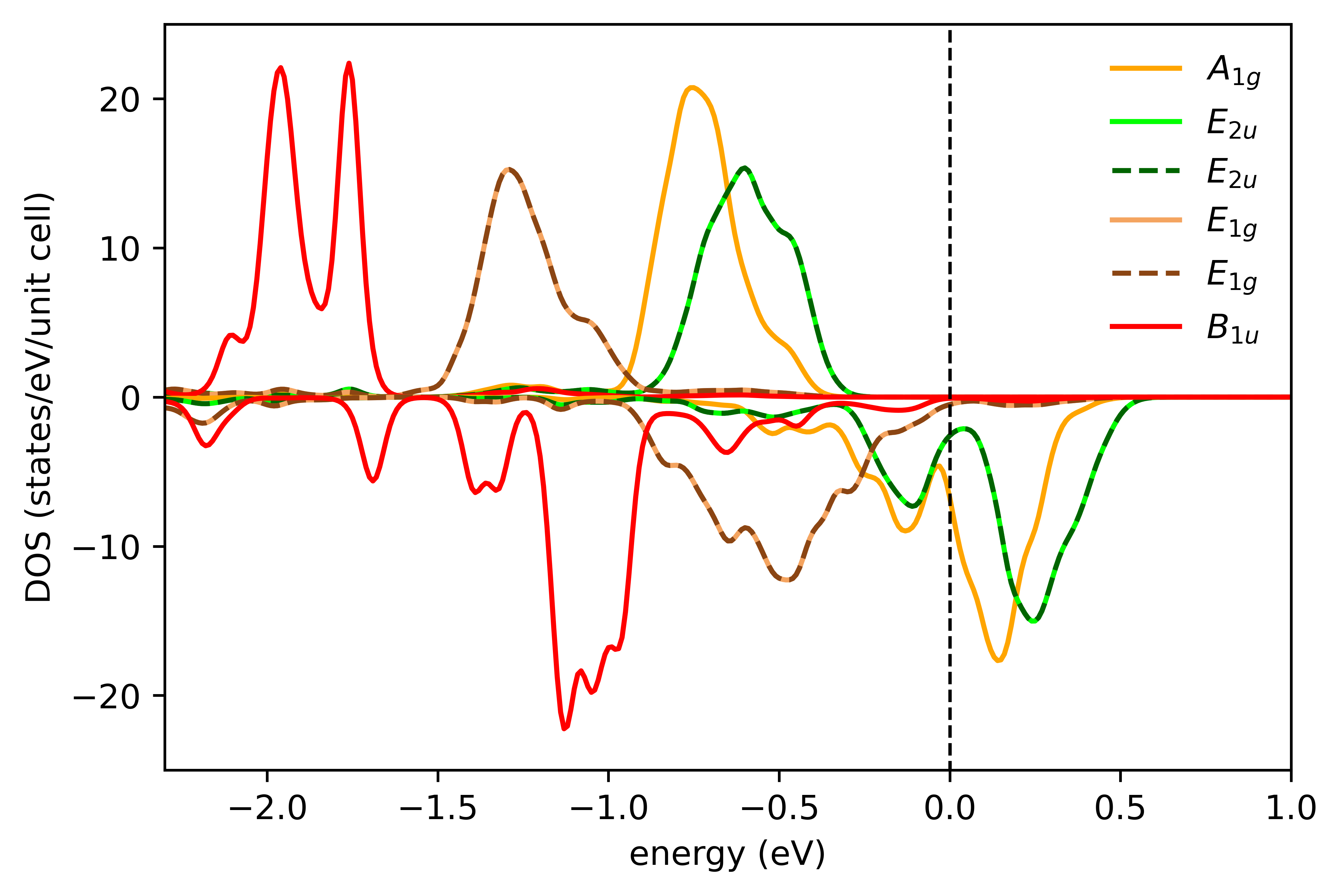}
  \caption{Spin-resolved density of states of ferromagnetic Na$_2$IrO$_3$ projected onto the six quasi-molecular orbitals. Spin-up(down) density of states are plotted in the positive (negative) part of vertical axis.}
  \label{fig:proj}
\end{figure}

Here, we provide the results of density of states projected on the quasi-molecular orbital basis for Na$_2$IrO$_3$ in ferromagnetic phase. The definition of six quasi-molecular orbitals is given in the Table I of Ref. \cite{mazin20122} , and the construction of projection operators is described in Ref\cite{foyevtsova2013ab}. The overlaps are computed within the PAW framework and the results are shown in Fig.\ref{fig:proj}.  
The calculation is carried out on the DFT ground state calculated with U=3.8eV. A spin-flip gap $\sim$0.3eV is found. And distribution of quasi-molecular orbitals are indeed localized in energy, indicating formation of quasi-molecular orbitals in the electronic structure. 

\section{orbital momentum basis $|L_Z\rangle$\label{AppC}}
The convention of $|L_Z\rangle$ basis used in the main text is
\begin{equation}
\begin{aligned}
  |L_Z=0\rangle&=\frac{1}{\sqrt{3}}(|d_{yz}\rangle+|d_{xz}\rangle+|d_{xy}\rangle) \\
  |L_Z=\pm1\rangle&=\pm\frac{1}{\sqrt{3}}(|d_{yz}\rangle+e^{\pm i\frac{2\pi}{3}}|d_{xz}\rangle+e^{\mp i\frac{2\pi}{3}}|d_{xy}\rangle), \\
\end{aligned}
\end{equation} 
where the local coordinates $xyz$ of $t_{2g}$ orbitals is determined by the cubic setting of oxygen-octahedra and the quantization axis $Z$ of orbital angular momentum corresponds to the local [111] direction perpendicular to the honeycomb plane.

\bibliographystyle{./statto}

\end{document}